\documentclass[12pt]{article}
\usepackage{amsmath}
\usepackage{setspace}
\usepackage{graphicx}
\usepackage{subfig}
\usepackage{natbib}
\usepackage{indentfirst}
\usepackage{lscape}
\usepackage{url}
\usepackage{JASA_manu}
\newcommand{\bfY}{{\mathbf{Y}}}
\newcommand{\bfZ}{{\mathbf{Z}}}
\newcommand{\bfX}{{\mathbf{X}}}
\newcommand{\bfW}{{\mathbf{W}}}
\newcommand{\bfD}{{\mathbf{D}}}
\newcommand{\bfx}{{\mathbf{x}}}
\newcommand{\bfw}{{\mathbf{w}}}
\newcommand{\bfm}{{\mathbf{m}}}
\newcommand{\bfh}{{\mathbf{h}}}
\newcommand{\bfalpha}{{\boldsymbol\alpha}}
\newcommand{\bfbeta}{{\boldsymbol\beta}}
\newcommand{\bfpsi}{{\boldsymbol\psi}}
\newcommand{\bfSigma}{{\boldsymbol\Sigma}}
\newcommand{\bfOmega}{{\boldsymbol\Omega}}
\newcommand{\bflambda}{{\boldsymbol\lambda}}
\newcommand{\bfxi}{{\boldsymbol\xi}}
\newcommand{\bfepsilon}{{\boldsymbol\epsilon}}
\newcommand{\bfeta}{{\boldsymbol\eta}}

\newcommand{\E}{{\mbox{E}}}

\newcommand{\Var}{{\mbox{Var}}}

\newcommand{\Cov}{{\mbox{Cov}}}

\begin{document}
\title{Using a Birth-Death Process to Account for Reporting Errors in Longitudinal Self-reported Counts of Behavior}
\author{Jihey Lee \\
John Wayne Cancer Institute \\
 2200 Santa Monica Blvd, Santa Monica, CA 90404
\\ Robert E. Weiss \\
Department of Biostatistics, Fielding School of Public Health,\\
University of California, Los Angeles, CA 90095-1772 \\
Marc A. Suchard \\
Departments of Biomathematics and Human Genetics,\\ David Geffen School of Medicine at UCLA, \\
Los Angeles, CA 90095-1766, USA and \\
Department of Biostatistics, UCLA Fielding School of Public Health, \\
Los Angeles, CA 90095-1772, USA
}
\maketitle
\newpage
\mbox{}
\vspace*{2in}
\begin{center}
\textbf{Author's Footnote:}
\end{center}
Jihey Lee is Senior Statistician, John Wayne Cancer Institute, 2200 Santa Monica Blvd, Santa Monica, CA 90404 (e-mail: jlee.sophia@gmail.com); Robert E. Weiss is Professor, Department of Biostatistics, School of Public Health, University of California, Los Angeles, CA 90095-1772 (e-mail: robweiss@ucla.edu); and Marc A. Suchard is Professor, Departments of Biomathematics and Human Genetics,\\ David Geffen School of Medicine at UCLA, Los Angeles, CA 90095-1766, USA and Department of Biostatistics, UCLA Fielding School of Public Health, Los Angeles, CA 90095, USA (e-mail: msuchard@ucla.edu). The authors thank Dr.\ Mary Jane Rotheram-Borus of the UCLA Center for Community Health for permission to use the CLEAR data.  The CLEAR study was supported by NIDA grant DA07903 and Weiss was partially supported by the Center for HIV Identification, Prevention, and Treatment (CHIPTS) NIMH grant MH58107; the UCLA Center for AIDS Research (CFAR) grant 5P30AI028697, Core H. MAS was supported by National Institutes of Health grant R01 AI107034 and National Science Foundation grant DMS 1264153. The authors thank Dr.\ Pamina M. Gorbach for valuable comments.
\newpage

\begin{center}
\textbf{Abstract}
\end{center}
We analyze longitudinal self-reported counts of sexual partners from youth living with HIV.  In self-reported survey data, subjects recall counts of events or behaviors such as the number of sexual partners or the number of drug uses in the past three months.  Subjects with small counts may report the exact number, whereas subjects with large counts may have difficulty recalling the exact number.  Thus, self-reported counts are noisy, and mis-reporting induces errors in the count variable.  As a naive method for analyzing self-reported counts, the Poisson random effects model treats the observed counts as true counts and reporting errors in the outcome variable are ignored.  Inferences are therefore based on incorrect information and may lead to conclusions unsupported by the data.  We describe a Bayesian model for analyzing longitudinal self-reported count data that formally accounts for reporting error.  We model reported counts conditional on underlying true counts using a linear birth-death process and use a Poisson random effects model to model the underlying true counts.  A regression version of our model can identify characteristics of subjects with greater or lesser reporting error.  We demonstrate several approaches to prior specification.

\vspace*{.3in}

\newpage
\section{INTRODUCTION}
\label{section:intro}

Self-reported count data often appear in public health studies; for example, the count of the number of cigarettes smoked in the past week (\citealt*{wan08}), the number of unprotected sex acts in the past four months (\citealt*{pat03}), and frequency of marijuana use in the last week (\citealt{pen89}).  In this paper, we analyze self-reported counts of sexual partners from the Choosing Life: Empowerment, Action, Results (CLEAR) longitudinal three-arm randomized intervention study designed to reduce HIV transmission and improve quality of life among HIV-infected youth.  Subjects were randomized equally to control or to one of two intervention delivery methods:  telephone and in-person.  Interest lies in comparing the two intervention delivery modes, comparing treatments to control, and in estimating effects of predictors known to be important. Our outcome in this paper is the self-reported number of sexual partners during the past three months, an important measure of sexual risk behavior (\citealt{rot01,lig05}).

Behavioral research on sexually transmitted diseases mostly depends on self-reports of sexual behavior (\citealt{jam04,fen01,cat90a}).  However, it has been argued that self-reports of sexual behaviors are not accurate and noisy for several reasons (\citealt*{kau91}).  Having zero or one partner is likely to be reported accurately, but reports of large numbers of partners are likely to be inaccurate, although the reports would still be large. The accuracy of self-reported sexual behavior has been found to be related to the number of sexual partners (\citealt{jam04}), the duration of recall periods (\citealt{cat90b}; \citealt*{kau91}), and one's propensity to engage in casual sex (\citealt{jam04}). Much research has aimed at improving the accuracy of self-reports of sexual behaviors (\citealt{tou97}).

The Poisson distribution is a frequent starting point for modeling counts of sex partners because it is a discrete probability distribution that takes on non-negative integer values. However, the Poisson distribution assumes equal mean and variance, and does not allow for over-dispersion when the variance of the counts is larger than the mean. An improvement is the Poisson random effects model (PREM). The PREM incorporates additional subject-specific coefficients allowing subject means to deviate from population means. Thus, the PREM accounts for unobserved heterogeneity among subjects, and injects more variation than the standard Poisson model. Further, the random effects induce correlation across longitudinal observations on a subject.

\citet*{gho09} extend a PREM longitudinal approach to a joint model accommodating various complications in self-reported counts of sexual events, but do not discuss errors in self-reports. \citet*{fad00} and \citet*{yan10} develop models for underreported counts. \citet*{bol97} extend a probit model accounting for over- and under-reporting error in the univariate response variable. \citet*{hei90}, \citet*{wan08}, and \citet*{hin08} propose methods for accommodating data reported or measured with error; they introduce a latent true count distinct from the observed count as do we.

In this paper, we model observed counts $Y_{ij}$ on subject $i$ at time $t_{ij}$ given underlying true but unobserved counts $Z_{ij}$ using a linear birth-death (BD) process, and model the true counts $Z_{ij}$ using a PREM.  Stochastic processes including the BD process have been used in many fields (for example, \citealt*{wil65,was80,lee94,dur99,mod01,van07,liu07}).  Our model exploits the BD process strictly as a sampling model for $Y_{ij}|Z_{ij}$ that has greater flexibility than the PREM. Further, our model differs from traditional measurement error or misclassification models in that our model accounts for errors in the outcome variable whereas measurement error models traditionally account for errors in covariates (for example, \citealt*{che79,sel86,whi88,del95,hen03}). The BD model is exciting because it provides a sampling distribution on the integers and it eases interpretability with variance parameters that are easily interpreted.

This article is organized as follows.  In section \ref{modelspec}, we discuss the PREM and present our new BD methodology for handling reporting errors.  In section \ref{bayesianinf}, we discuss Bayesian inference including priors and posterior distributions, and in section \ref{longmodel}, the PREM and our proposed models are applied to longitudinal self-reported count data from CLEAR. Section \ref{simulation} presents a simulation study comparing our BD model to the PREM.

\section{Model Specification}
\label{modelspec}

\subsection{Notation}

Let $Y_{ij}$ be the reported count for subject $i = 1, \ldots, n$ and observation $j = 1, \ldots, n_{i}$ at time $t_{ij}$ and let $Z_{ij}$ be the corresponding unobserved true count.  Each subject has a $p \times 1$ covariate vector $\bfx_{ij}$ measured at time $t_{ij}$ and define the $n_{i} \times p$ covariate matrix $\bfX_{i} = (\bfx_{i1}, \ldots, \bfx_{in_{i}})^{'}$, vector of responses $\bfY_{i} = (Y_{i1}, \ldots, Y_{in_{i}})^{'}$, and vector of unobserved true counts $\bfZ_{i} = (Z_{i1}, \ldots, Z_{in_i})^{'}$.

\subsection{Poisson Random Effects Model for the Unobserved True Count}

We model the unobserved true counts $Z_{ij}$ using a PREM.  We assume that the $Z_{ij}$ are independent conditional on a $p \times 1$ vector of fixed effects coefficients $\bfalpha = (\alpha_{1}, \ldots ,\alpha_{p})^{'}$ and an $r \times 1$ vector of random effects $\bfbeta_{i}$ multiplying by $\bfh_{ij}$, an $r \times 1$ vector of known predictors.  With a log link, we have
\begin{align}
Z_{ij} \mid \mu_{ij} & \sim  \mbox{Poisson} (\mu_{ij}), \nonumber\\
\mu_{ij} & \equiv E( Z_{ij} \mid  \bfalpha, \bfbeta_i),    \label{eq:PREM} \\
\log(\mu_{ij}) & = \mathbf{x}_{ij}^{'}\bfalpha + \bfh_{ij}^{'}\bfbeta_{i}, \nonumber
\intertext{and} \bfbeta_{i} \mid \mathbf{D}_{\beta} & \sim N_{r}(\mathbf{0},\mathbf{D}_{\beta}), \nonumber
\end{align}
where $\mu_{ij}$ is the mean of the $i$th subject's $j$th observation and $N_{r}(\mathbf{0}, \mathbf{D}_{\beta})$ denotes an $r$-dimensional multivariate normal random variable with mean $\mathbf{0}$ and $r \times r$ covariance matrix $\mathbf{D}_{\beta}$. The unconditional mean of $Z_{ij}$ is then $\nu_{ij} = \exp(\mathbf{x}_{ij}^{'}\bfalpha + \bfh_{ij}^{'}\mathbf{D}_{\beta} \bfh_{ij}/2)$.

\subsection{Linear Birth-Death Process for the Reported Counts Given the True Counts}

We conceptualize a reported value $Y_{ij}$ as the realization of a stochastic process beginning at the underlying true value $Z_{ij}$.  Specifically, we use a linear BD process $\{S(\tau)\}$ to model the conditional distribution of $Y_{ij}\mid Z_{ij}, \lambda_{ij}$, where $S(\tau) \in \{0,1,2,\ldots\}$ is an integer count over a conceptualized time interval $0 \leq \tau \leq 1$ with initial state $S(0)=Z_{ij}$ and final state $S(1) = Y_{ij}$, and $\lambda_{ij}$ parameterizes the linear BD process.  We denote this distribution
\begin{align}
Y_{ij}\mid Z_{ij}, \lambda_{ij} \sim \mbox{BD}(Z_{ij}, \lambda_{ij}). \label{eq:BDprocess}
\end{align}
Traditionally, for a stochastic process $\tau$ is real time; in our model, however, $\tau$ is not an actual time but merely indexes the stochastic process $\{S(\tau)\}$ of which we only make use of the distribution at $\tau=1$.  A traditional linear BD process has two parameters, a per-capita birth rate $\lambda_{B,ij}$ and a per-capita death rate $\lambda_{D,ij}$.  The process assumes that as $\tau$ increases from 0, $S(\tau)$ increases or decreases by 1 with instantaneous birth rate $\lambda_{B,ij} S(\tau)$ and death rate $\lambda_{D,ij} S(\tau)$ at time $\tau$. The process has an absorbing state at $S(\tau) = 0$.  When used as a sampling density for $Y_{ij}$, the birth rate $\lambda_{B,ij}$ and death rate $\lambda_{D,ij}$ can be interpreted as an individual's propensity to over- and under-report, respectively.

To simplify, we assume $\lambda_{B,ij} = \lambda_{D,ij} \equiv \lambda_{ij}$ which leads the mean of the reporting distribution to be unbiased for the underlying true value
\begin{align}
\E(Y_{ij}\mid Z_{ij}, \lambda_{ij}) = Z_{ij} \nonumber
\end{align}
and the conditional variance is proportional to both $Z_{ij}$ and $\lambda_{ij}$
\begin{align}
\Var(Y_{ij} \mid Z_{ij}, \lambda_{ij}) = 2\lambda_{ij} Z_{ij} \nonumber
\end{align}
(\citealt*{nor64}). The variance of the observed counts $Y_{ij}$ is then
\begin{align}
\Var(Y_{ij}) =  (2*\lambda_{ij}+1)\nu_{ij} + \nu_{ij}^2(\exp(\bfh_{ij}' \mathbf{D}_{\beta} \bfh_{ij}) -1) \label{eq:varvar}
\end{align}
and the covariance between $Y_{ij}$ and $Y_{ik}$, $j \ne k$ is
\begin{align}
\Cov(Y_{ij},Y_{ik}) =  \nu_{ij}\nu_{ik} (\exp(\bfh_{ij}' \mathbf{D}_{\beta} \bfh_{ik}) -1) \label{eq:cov}
\end{align}
which follow from standard rules of conditional probability and results in \citealt{aitchison1989multivariate}. The variance \eqref{eq:varvar} is increased by $2*\lambda_{ij}\nu_{ij}$ over that of the standard PREM model, while the covariance \eqref{eq:cov} is unchanged from the PREM model.

Observations with large $Z_{ij}$ and/or $\lambda_{ij}$ lead to large variances of $Y_{ij}$, and are associated with low recall accuracy.  The BD rate $\lambda_{ij}$ represents the relative accuracy of reports or recall.  If an individual mis-reports the number of events, then $Y_{ij}$ would be greater or less than $Z_{ij}$, and were $Z_{ij}$ known, the difference $(Y_{ij}-Z_{ij})$ would be a type of residual and is a measure of the accuracy of observation $Y_{ij}$.

A derivation of the sampling  density $p(Y_{ij} \mid Z_{ij}, \lambda_{ij})$ is given in Appendix \ref{BD prob}.  Figure~\ref{fig:Ydistri} illustrates example distributions of reported counts $Y \mid Z, \lambda \sim \mbox{BD}(Z,\lambda)$ for 9 combinations of $Z$ and $\lambda$.  Row 1 has small $\lambda=0.5$ indicating relatively accurate reports, row 2 reports $\lambda=3$, and row 3 demonstrates $\lambda=7$ for relatively inaccurate reports.  Column 1 has $Z=2$, column 2 has $Z=15$, and column 3 has $Z=50$ for a modest, medium, and large number of underlying counts.  Figures~\ref{fig:Ydistri}\subref{Ydistri_2}, \ref{fig:Ydistri}\subref{Ydistri_3}, and \ref{fig:Ydistri}\subref{Ydistri_6} demonstrate modes not at the true count but at zero due to the absorption of the process $S(\tau)$ at zero when $\lambda$ is large compared to $Z$.  The others return modes at $Z$.

\subsection{A Log-linear Regression Model for the Birth/Death Rate}

The simplest model allows subjects to share a common BD rate $\lambda_{ij} \equiv \lambda$, but it seems unrealistic to assume all subjects have the same propensity to mis-report.  We expect $\lambda_{ij}$ to vary across subjects and even within subject over time depending on time-fixed and time-varying covariates.  Because $\lambda_{ij}>0$, we use a log-linear regression model for $\lambda_{ij}$
\begin{align}
\lambda_{ij} = \exp(\bfw_{ij}^{'}) \bfpsi \label{eqloglinear}
\end{align}
where $\bfw_{ij}$ denotes a $q \times 1$ covariate vector for the $i$th subject at time $t_{ij}$, $\bfpsi = (\psi_{1},\ldots,\psi_{q})^{'}$ is a vector of regression coefficients for the fixed effects.
We call \eqref{eqloglinear} the BD model for short.

\section{Bayesian Inference}
\label{bayesianinf}

\subsection{Prior Distributions}

We specify the priors for parameters $\bfalpha$, $\bfpsi$, and $\mathbf{D}_{\beta}$, to be independent \textit{a priori}.  For the fixed effects, we assume a traditional normal prior: $\bfalpha \sim N(\bfm_{\alpha}, \bfSigma_{\alpha})$ and $\bfpsi \sim N(\bfm_{\psi},\bfSigma_{\psi})$, where most commonly $\bfSigma_{\alpha}$ and $\bfSigma_{\psi}$ are diagonal matrices with known diagonal elements.  For the covariance matrix $\mathbf{D}_{\beta}$, we assume $\mathbf{D}_{\beta} \sim \mbox{IW}_{r}(\bfOmega_{\beta},m_{\beta})$, where $\mbox{IW}_{r}(\bfOmega_{\beta},m_{\beta})$ denotes an $r\times r$ inverse-Wishart distribution with degrees of freedom (df) $m_{\beta} \geq r$ and mean $\bfOmega_{\beta}/(m_{\beta}-r-1)$.

We consider several approaches to the problem of specifying $\bfm_{\alpha}$, $\bfSigma_{\alpha}$, $\bfm_{\psi},$ and $\bfSigma_{\psi}$ for this model: (1) an approach based on previous studies (PS) reported in the literature, (2) a pure elicitation (PE) approach, (3) data augmentation (DA) and (4) analysis of a previous similar data set (DS).  These approaches are not necessarily disjoint; the methods can be mixed and we combine them opportunistically.

In the PS approach, $\bfm_{\alpha} = (m_{\alpha,k})$ are point estimates taken from papers in the literature, as are the standard errors $\Sigma_{\alpha, k}$ and $\bfSigma_{\alpha}$ is diagonal with $k$th diagonal element $\Sigma_{\alpha, k}$.  However, while we are often willing to generate prior estimates from the literature, we feel standard errors from the literature are usually over-precise for application to novel data.

A PE approach can be used for $\bfm_{\alpha}$ and the diagonal elements of $\bfSigma_{\alpha}$ using what we call the \textit{point and range method}.  Often we may specify a prior point estimate $m_{\alpha,k}$ of $\alpha_k$ and suppose we can state that we expect a subject with covariate $x_{ijk} = 1$ has on average at most $d$ times as many partners as a subject with
$x_{ijk} = 0$ and that $d$ is at the edge of a 95\% probability interval.  We find $\Sigma_{\alpha, k}$ by solving $\exp(m_{\alpha,{k}} + 1.96\Sigma_{\alpha, k}) = d$.  Choices for $\bfm_{\alpha}$ include the journal article estimate or $\bfm_{\alpha}=0$, to keep the prior neutral as to the sign of $\alpha_k$.  The value $d$ may be elicited as a number that is ``too big"; the resulting prior is appropriately centered and not overly informative while still being proper and sensibly prejudiced against \textit{a priori} ridiculous values of $\alpha_k$.

For a DA prior (\citealt*{bed96}, \citeyear{bed97}), we construct a prior data set $\mathbf{x}^{0}_{k_1}$ and $Z^{0}_{k_1,\mbox{\tiny PREM}}$ for $k_1 = 1, \ldots, K_1$ as $K_1$ prior representative cases for the PREM part and  $\bfw^{0}_{k_2}$, $Y^{0}_{k_2}$, and $Z^{0}_{k_2,\mbox{\tiny BD}}$ for $k_2 = 1, \ldots, K_2$ as $K_2$ prior representative cases for the BD part.  We then  plug this data into the likelihood to get a function proportional to the desired prior.  One might use either a fixed effects Poisson regression likelihood or a random effects regression likelihood for the DA prior in the PREM model, and we used the latter to be consistent with the PREM model.
The resulting DA prior distribution then becomes
\begin{eqnarray}
\lefteqn{p_{DA}(\bfalpha,\bfbeta^0,D_\beta,\bfpsi \mid \bfX^0, \bfY^0, \bfZ^0_{\mbox{\tiny PREM}}, \bfZ^0_{\mbox{\tiny BD}}, \bfW^0)}\label{eq:DAprior} \\
& & \;\propto \displaystyle \prod_{k_1=1}^{K_1}\left\{\frac{\exp\{Z^0_{k_1,\mbox{\tiny PREM}}(\bfx^{0'}_{k_1}\bfalpha + \beta_{k_1}^0)-\exp(\bfx^{0'}_{k_1}\bfalpha + \beta_{k_1}^0)\}}{Z^0_{k_1,\mbox{\tiny PREM}}!}\right\}  \nonumber\\
& & \;\times D_\beta^{-K_1/2}\exp \left(-\frac{1}{2D_\beta}\sum_{k_1=1}^{K_1}(\beta^0_{k_1})^2\right) \times
D_{\beta}^{-(a+1)}\exp(-b/D_{\beta}) \nonumber  \\
& & \;\times \displaystyle \prod_{k_2=1}^{K_2} p(Y_{k_2}^0 \mid Z^0_{k_2,\mbox{\tiny BD}}, \bfpsi). \nonumber
\end{eqnarray}
We introduce artificial $\beta_k^0$'s in \eqref{eq:DAprior}, however we do not care about them; the purpose is to produce a prior for $\bfalpha$ and $D_\beta$.  The pre-prior $D_\beta \sim IG(a,b)$ then guarantees that \eqref{eq:DAprior} is a proper prior as long as $K_1 \geq p$ and $K_2 \geq q$.
We may take the resulting density \eqref{eq:DAprior} as our prior or, for convenience, we may take the means and standard deviations (SDs) from analysis of this prior data set as the prior parameters for the data set of interest.

In the DS approach, we can use estimates and covariance matrices from the analysis of previous similar data sets as the prior parameters for the data set of interest.  One advantage is that covariances among the regression parameters can be brought into the covariance matrix.

Prior specification in Bayesian modeling requires substantial subject matter knowledge and we discuss details of these approaches in the specific context of our data set in section \ref{longmodel} where we also present results of our data analysis.

\subsection{Posterior Distribution}

Let $\bfbeta = (\bfbeta_{1}, \ldots, \bfbeta_{n})^{'}$, $\bflambda = (\lambda_{11}, \ldots, \lambda_{nn_{n}})^{'}$, $\bfZ = (\bfZ_{1}, \ldots, \bfZ_{n})$ and let $N \times q$ matrix $\bfW = (\bfw_{11}, \ldots, \bfw_{nn_{n}})^{'}$ where $N = \sum_{i=1}^{n} n_{i}$ is the total number of observations.  The joint posterior distribution of $\bfZ, \bfalpha, \bfbeta, \bflambda, \bfpsi, \mathbf{D}_{\beta}$ is $p(\bfZ, \bfalpha, \bfbeta, \bflambda, \bfpsi, \mathbf{D}_{\beta}, \mid \bfY, \bfX, \bfW)$ and is given in Appendix \ref{PosteriorDisti}.

\subsection{Computing Overview}

The underlying true counts $Z_{ij}$ are discrete random variables taking values on the non-negative integers; the other unknowns are continuous.  The joint posterior distribution is intractable and we draw inference through sampling from the posterior using Markov chain Monte Carlo (MCMC) methods (\citealt{met53}; \citealt*{has70,gel90,car92,cha95,wal98,chi99}), specifically using Metropolis and Metropolis-Hastings (MH) steps within a random scan Gibbs sampling algorithm (\citealt*{rob97,rob04,liu08}).

For most steps, we consider adaptive auto-optimizing transition kernels (\citealt*{ros11}) that automatically adjust the scale of the proposal distribution as the MCMC runs in an attempt to achieve a specific acceptance probability $\pi$.  Let $\kappa_{m}$ be the scale parameter of a proposal distribution at iteration $m$, and let $\theta_{m}$ be the acceptance frequency for the proposal up to iteration $m$.  Then, we set scale $\kappa_{m+1}$ for iteration $(m+1)$ to
\begin{align}
\kappa_{m+1} = \kappa_{m} + \frac{\theta_{m} - \pi}{t(m) + 1},
\end{align}
where $t(m)$ is a monotonic transform of $m$, such as $t(m) = m$, or $t(m) = \sqrt{m}$.  The target acceptance probability $\pi$ can be set differently for different parameters if warranted.

\begin{center}
\section{\textsc{Longitudinal Modeling for Sex Partner Counts}}
\label{longmodel}
\end{center}

\subsection{Data from the CLEAR Study}

Our primary outcome measurement is the self-reported number of sexual partners in the past 3 months for $n = 175$ HIV+ young people.  Observations were taken at time 0, the baseline observation, and at 3, 6, 9, and 15 months.  Roughly 80\% of subjects are available at each follow-up time, suggesting that the data are at worst intermittently missing, and drop-out is not a major concern.

\subsection{Predictors}

For the PREM fixed effects, we include time-fixed indicators of injection drug use (IDU) (yes=1, no=0) and men who have sex with men (MSM) (yes=1, 0=women and also men who have sex with women only).  It is common to combine heterosexual men and women into a single category in these analyses (\citealt{bol06}).  We include two time-varying indicators, one for trading sex for money, drugs, food or housing in the past three months (yes = 1 for any trading, no=0) (TRADE) and one for engaging in casual sex in the past 3 months (yes=1, no=0) (CASUAL).  Subjects are randomized to one of three treatment groups, telephone delivery, in-person delivery and control.  All three groups are modeled as having the same average baseline number of partners.  At follow-up we include 12 indicators for the 3 intervention means at the 4 measurement times.  For the random effects, we take $r=1$ and $h_{ij}=1$ giving a random intercept model; $\beta_{i}$ and $D_{\beta}$ are then scalars.  We set $a=3$ and $b=2$ in the prior for $D_{\beta} \sim IG(a,b)$ to obtain a proper prior with mean and variance equal to 1.  

Our BD process is fundamentally a variance model; typically there is less information in data about variances than about means and we simplify our loglinear model for the birth rate parameter as compared with the mean.  We include TRADE and CASUAL as covariates.  At baseline all subjects are in a single group.  For all post-baseline times, we include three indicators for the three treatment groups.

\subsection{Prior Specification from Previous Studies}

\emph{A combined PS/PE Prior for PREM}.
We construct one prior using a combination of information from previous studies and from elicitation.
Prior means and SDs for the PREM are presented in the PS/PE prior columns of Table~{\ref{estimates}}.
We take the point estimates for MSM from \citet{sol09} and those for IDU and TRADE from \citet*{dil02}.  The prior mean for CASUAL is obtained from \citet{kie06} where the outcome variable is the number of unprotected (vaginal or anal) sex events per partner.  We presume the number of partners proportionally increases with the number of acts.

We assume subjects have one partner on average at baseline given no IDU, MSM, CASUAL, and TRADE, which gives $\log 1 = 0$ prior mean for the intercept.  We specify zero prior means for time effects and interactions between time and intervention groups because we have little prior knowledge about time trend and intervention effects and, for this prior, we wish to not directly input prior beliefs about the direction of treatment and time effects.

To specify the prior variance, we assume that MSM, IDU, TRADE, and CASUAL may have up to 15, 20, 30, and 30 times as many partners as non-MSM, -IDU, -CASUAL, and -TRADE, respectively at the outside of a 95\% prior interval.  For the prior variance of the intercept, we assume that at baseline 95\% of non-MSM, -IDU, -CASUAL, and -TRADE subjects have from 1/30 to 30 partners on average.  We set the prior variances for the time effects and interaction terms equal to 4 to represent vague prior knowledge.

\emph{PE Prior for the BD process}.
We assume that a 95\% prior interval of the birth rates is from 1/80 to 80 at baseline with CASUAL=TRADE=0, which gives a prior mean of 0 and prior SD 2.236 for the intercept.  We similarly specified prior means and ranges for the other variables, and the resulting means and SDs are shown in Table~{\ref{birth rate result}}, columns under Prior.

\subsection{Data Augmentation Prior}

\emph{DA for the PREM}.  We set a separate prior for the fixed effects with a DA prior and this time we do include proper informative prior information about the treatment groups.
We assume that when a subject is not an IDU or MSM, and has $\mbox{CASUAL} = 0$ and 
$\mbox{TRADE} = 0$, the subject has 1 partner on average at baseline which is close to the 0.7 in the CLEAR data.  We assume that there are no changes in the number of partners at 3, 6, 9, and 15 months from the baseline in the control group, but in the telephone and in-person intervention groups the number of partners are reduced to 0.8 and 0.5 times at follow-up months compared to the baseline; we expect the in-person intervention to be more effective.  We also assume that IDU and MSM subjects have twice as many partners as non-IDU and non-MSM subjects, and subjects participating in casual sex and trade have 4 and 8 times, respectively, as many partners as subjects not engaging in such acts.  The prior data are shown in Table~{\ref{DA for PREM}}.

\emph{DA for the BD process}.
We estimate a priori a 2 or 3 partner difference between observed and true counts at baseline and at follow-ups in the control group when $\mbox{TRADE}=0$ and $\mbox{CASUAL}=0$.  At follow-up, telephone and in-person intervention groups are assumed to have three and zero difference from control group, respectively because we assume subjects who received in-person intervention sessions would pay more attention to their behaviors.  We assume that subjects involved in casual sex and trade report a number of partners further from the true count.  These prior data are shown in Table~{\ref{DA for PREM}}.

We combine this prior data with CLEAR data and proceed through a Bayesian inference without pre-priors on the model parameters except for the $D_\beta$ pre-prior that is needed to make a proper prior.

\subsection{Prior Based on Previous Data Sets}

Teens Linked to Care (TLC) was a close predecessor study to CLEAR enrolling 308 HIV+ youth and completed prior to CLEAR. Since CLEAR was a second generation version of TLC, the two studies share many similarities: goals, target populations, and geographic areas where subjects resided.  Similar measurements were taken at baseline and re-evaluated at 3, 6, 9, and 15 months in both studies.  The main differences are (i) participants were recruited from 1991 to 1996 in TLC and from 1999 to 2000 for CLEAR, and (ii) TLC randomized subjects 50-50 to in-person intervention or control while CLEAR had two interventions plus control.

We analyze TLC with a vague proper prior for the regression parameters and $D_\beta \sim IG(3,2)$.  We take the resulting posterior means $\bar{\bfalpha}_{\mbox{\tiny TLC}}$, $\bar{\bfpsi}_{\mbox{\tiny TLC}}$ and posterior variances $\bfSigma_{\alpha,\mbox{\tiny TLC}}$, $\bfSigma_{\psi,\mbox{\tiny TLC}}$ as the prior means $\bfm_\alpha = \bar{\bfalpha}_{\mbox{\tiny TLC}}$, $\bfm_\psi = \bar{\bfpsi}_{\mbox{\tiny TLC}}$ and prior variances $\bfSigma_\alpha = g\bfSigma_{\alpha,\mbox{\tiny TLC}}$, $\bfSigma_\psi = g\bfSigma_{\psi,\mbox{\tiny TLC}}$ for CLEAR.  The constant $g$ multiplies $\bfSigma_{\alpha,\mbox{\tiny TLC}}$ and $\bfSigma_{\psi,\mbox{\tiny TLC}}$ to inflate variances and reduce the prior contribution to the analysis. We take $g=34.46 (\approx 1034/30)$ in our CLEAR analysis assuming the 1034 observations in the TLC prior data are worth 30 observations in the CLEAR analysis. Prior means and SDs obtained from analyzing the TLC data are presented in the DS prior columns of Table~{\ref{estimates}}.
To deal with the different numbers of interventions in the two data sets, let $\bfalpha_{\mbox{\tiny T}}$ denote the $4\times 1$ vector for the 4 interactions between intervention and follow-up in TLC, and let $\bfalpha_{\mbox{\tiny CI}}$ and $\bfalpha_{\mbox{\tiny CT}}$ be the $4\times 1$ vectors for the 4 interactions between the in-person/telephone intervention group and follow-ups in CLEAR.  In this prior specification, we assume a priori intervention effects in TLC are the average of the 2 intervention effects in CLEAR; $\bfalpha_{\mbox{\tiny T}}=(\bfalpha_{\mbox{\tiny CI}}+\bfalpha_{\mbox{\tiny CT}})/2$.

We specify a normal prior with zero prior mean and compound symmetry prior covariance with correlation 0.5 for the difference of the intervention effects $(\bfalpha_{\mbox{\tiny CI}}-\bfalpha_{\mbox{\tiny CT}})$.  To specify the prior variance, we assume that either intervention group might have up to 10 times as many partners as the other intervention group at each follow-up time at the outside of a 95\% prior interval when everything else is controlled for giving a prior SD of $(\log 10)/1.96 = 1.175$.

We use the same procedure for the BD parameters $\bfpsi$ except that the intervention effect is a scalar rather than a vector at follow-up.  Let $\psi_{\mbox{\tiny T}}$ denote the interaction between intervention and post-baseline in TLC, and let $\psi_{\mbox{\tiny CI}}$ and $\psi_{\mbox{\tiny CT}}$ the interactions between in-person/telephone intervention group and post-baseline in CLEAR. We assume $\psi_T=(\psi_{\mbox{\tiny CI}}+\psi_{\mbox{\tiny CT}})/2$ and $(\psi_{\mbox{\tiny CI}}-\psi_{\mbox{\tiny CT}}) \sim N(0,1.175^2)$.

We also carry along prior information for $D_\beta$.  We let $D_\beta \sim IG(a,b)$ for the prior in CLEAR.  Assuming that TLC prior data are worth 30 observations and each subject has 5 observations, we arrive at a prior sample size of 6 which gives $a=6/2=3$.  The scale parameter $b$ is determined by solving $\bar{D}_{\beta,\mbox{\tiny TLC}} = b/(a-1)$ giving a value $b=0.549$.

\subsection{Computational Details}

The fixed effects parameters $\bfalpha$ are separated into coefficients of time-varying $(V)$ and time-fixed $(F)$ coefficients $\bfalpha = (\bfalpha^{(F)'}, \bfalpha^{(V)'})'$ and are updated in separate MH steps.  In our random scan Gibbs sampling, probabilities for selecting updates are set to be 0.2 for each of $\bfbeta$ and $\bfZ$, 0.26 for $\bfalpha^{(V)}$, 0.07 for each of $D_\beta$, $\bfalpha^{(F)}$, and $\bfpsi$, and 0.13 for $\bflambda$.  Larger probabilities are given to parameters with poorer convergence to improve efficiency.  We use $t(i) = \sqrt{i}$ in our adaptive auto-optimization algorithm.

Of the 10,010,000 MCMC samples we generate, the first 10,000 samples are discarded as burn-in, and of the next 10,000,000 samples, we save every 100th sample.  Code is implemented in Java.  Sampling details for all parameters are given in Appendix \ref{sampling algorithms}. Convergence as investigated through time series plots and autocorrelation plots seemed satisfactory.

\subsection{Results}

\begin{figure}
\includegraphics[scale=.6]{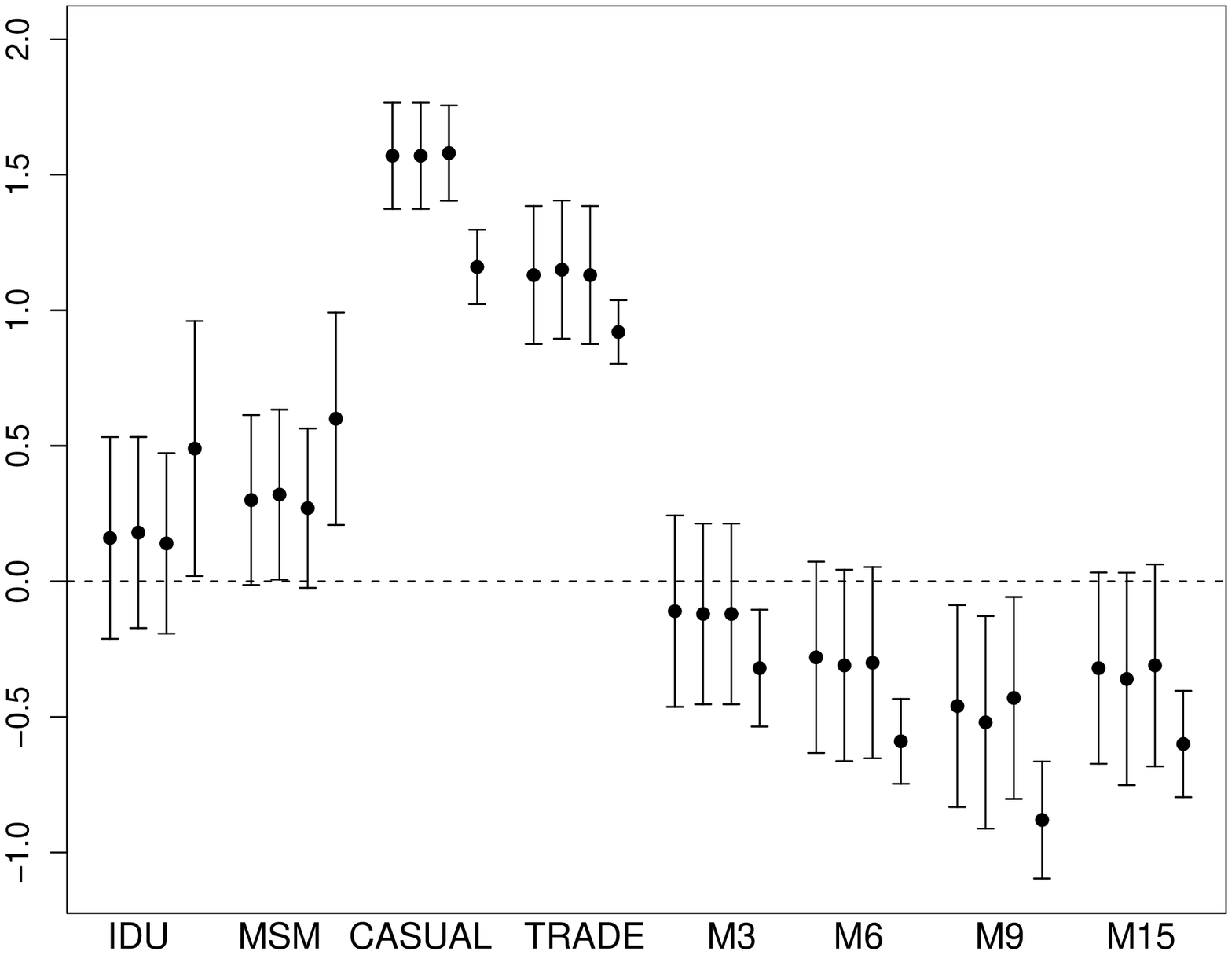}
\includegraphics[scale=.6]{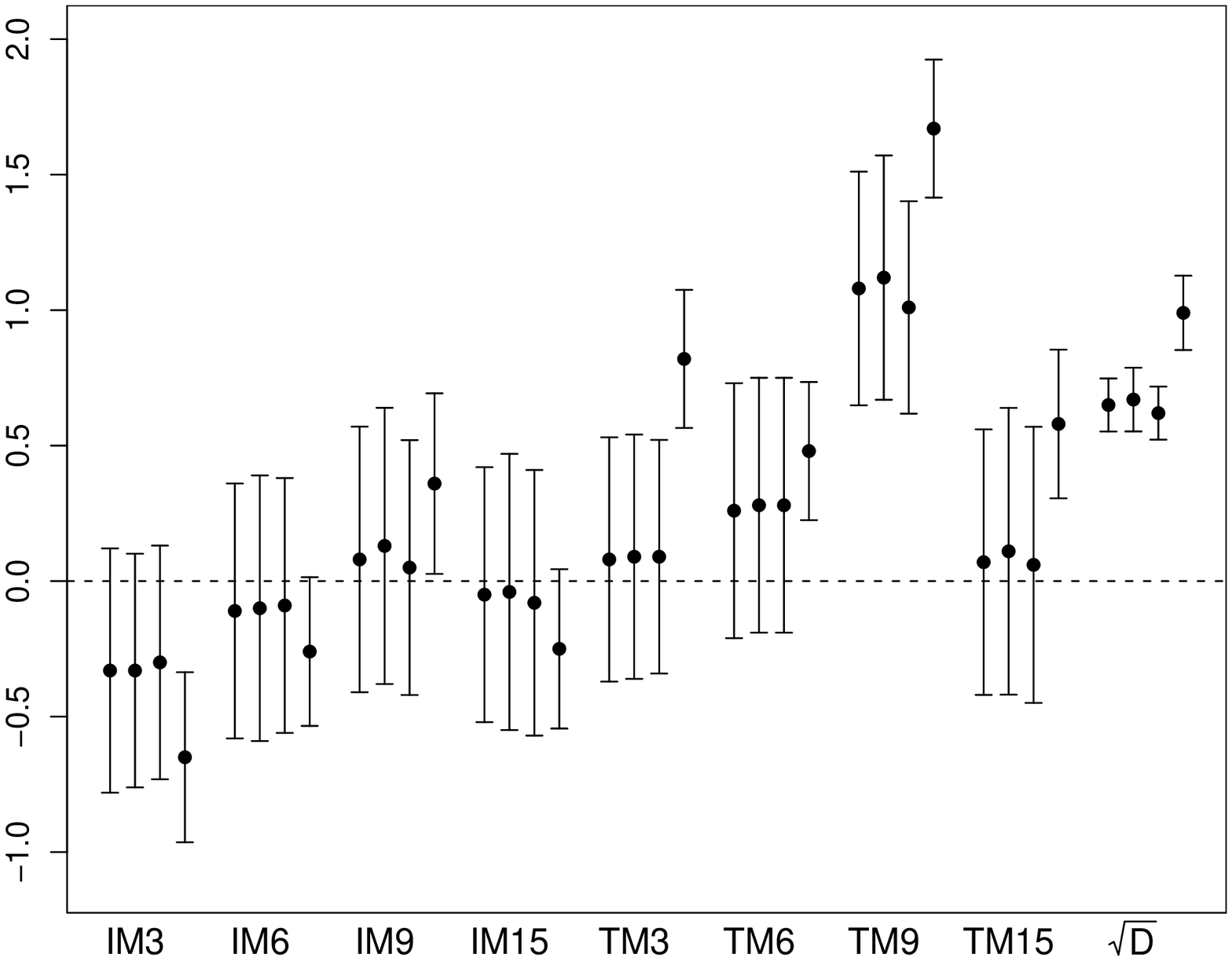}
\caption{\label{figtable1} Bayesian 95\% credible intervals (CIs) for the predictors and square root of the random intercept variance $D$ in the Poisson random effects model (PREM) component.  Each predictor has 4 CIs; the first 3 CIs are from the birth-death PREM (BDPREM) with the previous study/pure elicitation (PS/PE), data augmentation (DA) prior, and previous data set (DS) prior and the last one is from the PREM with PS/PE prior. Coefficients M3, M6, M9, and M15 are indicators for the control group for the follow-up months 3, 6, 9, and 15; IM3, IM6, IM9, and IM15 and TM3, TM6, TM9, and TM15 are interactions between in-person and telephone intervention group and the 4 follow-up months.}
\end{figure}

\begin{figure}
\centering
\includegraphics[scale=.8]{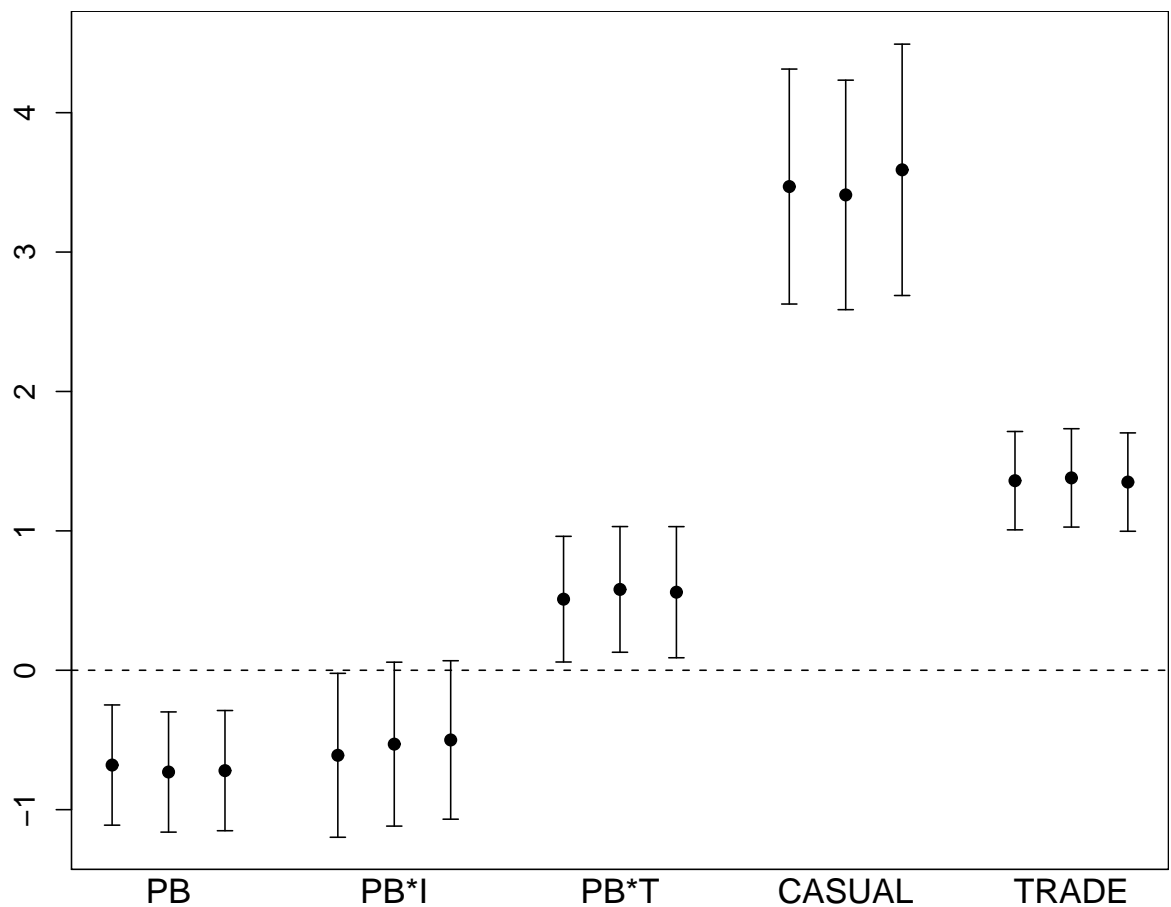}
\caption{\label{figtable2} Bayesian 95\% CIs for the predictors in the BD process from the BDPREM.  Each predictor has 3 CIs from analysis with the PS/PE, DA, and DS prior.  PB is an indicator for the measurements taken at post-baseline, PB*I and PB*T are interactions between PB and in-person/telephone intervention group.}
\end{figure}

We call our new model \eqref{eq:PREM}, \eqref{eq:BDprocess}, and \eqref{eqloglinear} the BDPREM.  We fit the BDPREM with all three priors and compare them to the PREM with combination PS/PE prior.  Posterior means and SDs for all four prior-model combinations are presented in Table~{\ref{estimates}}.  Table~{\ref{birth rate result}} presents results for the predictors of the BD process.  Figure~\ref{figtable1} plots posterior means and 95\% posterior intervals for the regression coefficients and $D^{1/2}$ for the four prior-model combinations and Figure~\ref{figtable2} plots similarly for the BD process component with three priors. The results for the BDPREM are similar across the three priors.

To compare model fits we calculate log marginal likelihoods for the PREM and BDPREM under the PS/PE prior, which are $-2170.41$ and $-1575.16$, respectively using Chib's method (\citealt*{chi01}) giving an enormous Bayes factor of $\exp(500)$ in favor of the BDPREM.

For the BD process component of the model, the telephone treatment group reports are noisier than baseline reports which are noisier in turn than the control and in-person treatment groups.  CASUAL and TRADE behaviors are associated with substantially increased reporting error.

\begin{landscape}
\begin{figure}
\centering
\subfloat[BDPREM]{
\includegraphics[scale=.55]{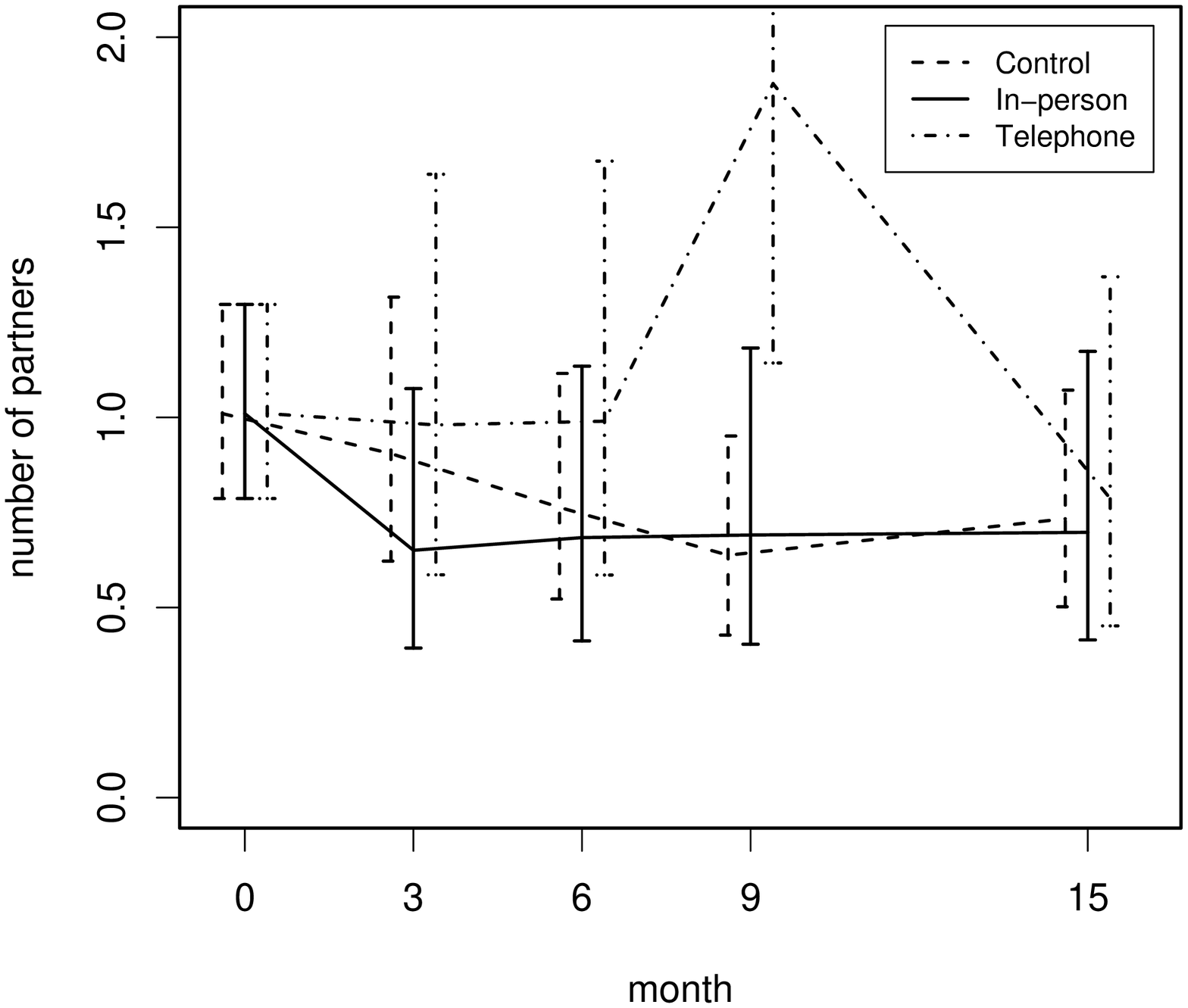}
\label{tmtBDPREM}
}
\subfloat[PREM]{
\includegraphics[scale=.55]{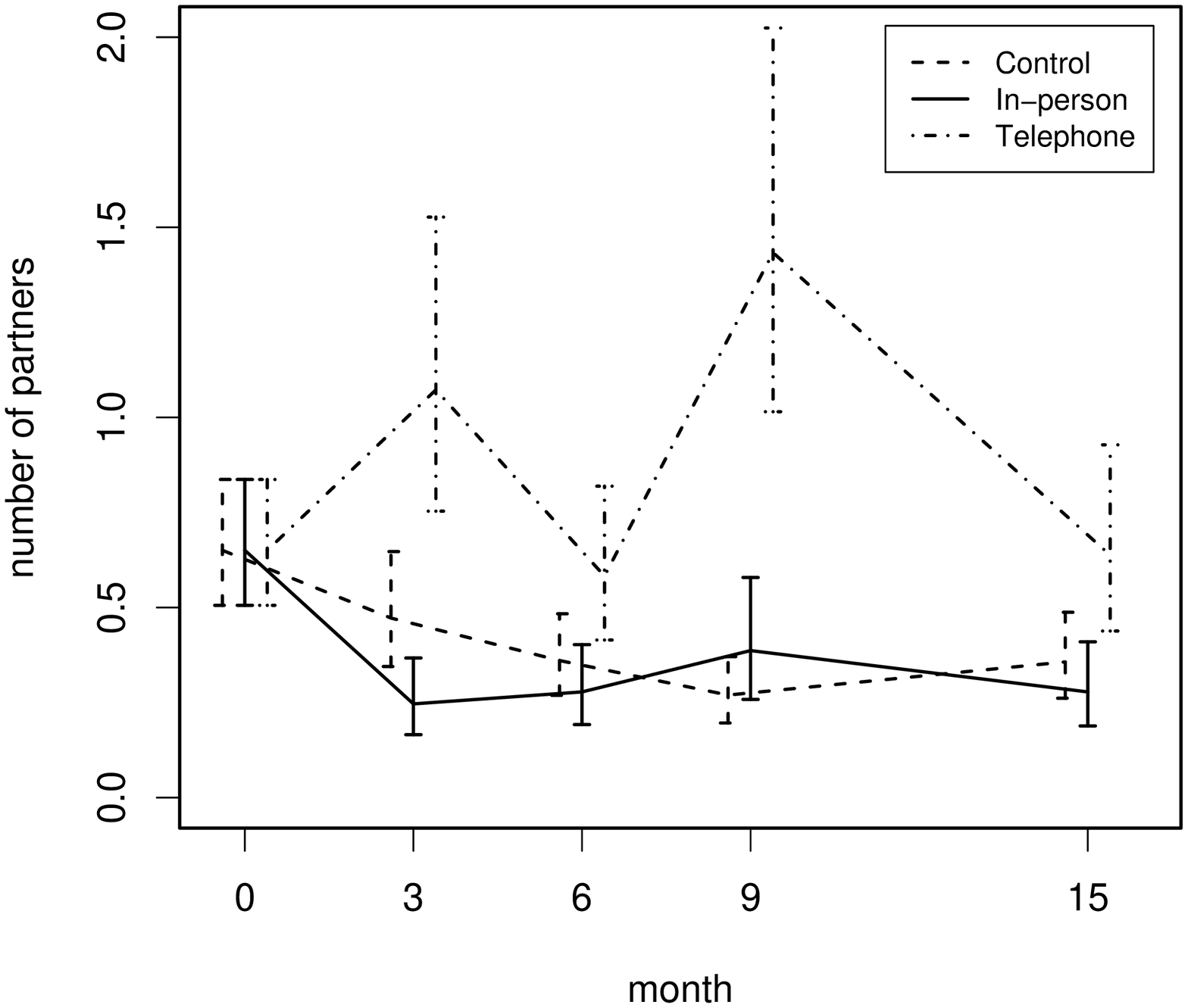}
\label{tmtPREM}
}
\caption{\label{fig:tmteffect}Prediction plots of the average number of partners for each intervention group for subjects with MSM=1 and IDU=CASUAL=TRADE=0 using (a) the BDPREM and (b) the PREM.  We show 95\% prediction intervals at each follow-up month. We set the same Y axis in (a) and (b) for comparability.  The upper limit for the telephone group at month 9 in (a) is 3.08.}
\end{figure}
\end{landscape}

All intervention effects are attenuated in the BDPREM compared to the PREM with smaller absolute regression coefficients for follow-up times and interactions between in-person/telephone intervention and follow-up times, and greater SDs.
Figure~\ref{fig:tmteffect} illustrates time trends for the 3 intervention groups resulting from (a) the BDPREM and (b) the PREM both with PS/PE prior, $\beta_i=0$ and given MSM=1 and IDU=CASUAL=TRADE=0, the largest subpopulation in the CLEAR data set.
Figure~\ref{fig:tmteffect}\subref{tmtBDPREM} demonstrates that 95\% prediction intervals for the 3 groups overlap at all time points except for 9 month telephone intervention group, implying generally similar trends in numbers of partners among the 3 intervention groups.  The substantial difference in the telephone group at month 9 results from 3 subjects reporting far greater numbers of partners at month 9 than at other months.  When we re-fit the BDPREM after excluding those subjects, the 9 month telephone group effect is no longer significantly different from the other group.  We define a parameter or contrast in a Bayesian analysis as significantly different from zero when a 95\% posterior interval for the parameter does not contain zero.

In contrast, in the PREM results presented in Figure~\ref{fig:tmteffect}\subref{tmtPREM}, the telephone intervention group shows significantly higher numbers of partners and shows a different trend than the in-person and control groups at all follow-up times.

In the PREM, IDU and MSM are significantly associated with having more partners, but the association does not retain significance under the better fitting BDPREM. These differences have important public health implications. CASUAL and TRADE are associated with increased partners in both models with stronger effects in the BDPREM.

\begin{figure}
\centering
\subfloat[$\bar{Z}_{64,2} = 3.87, \bar{\lambda} = 0.02$]{
\includegraphics[scale=.35]{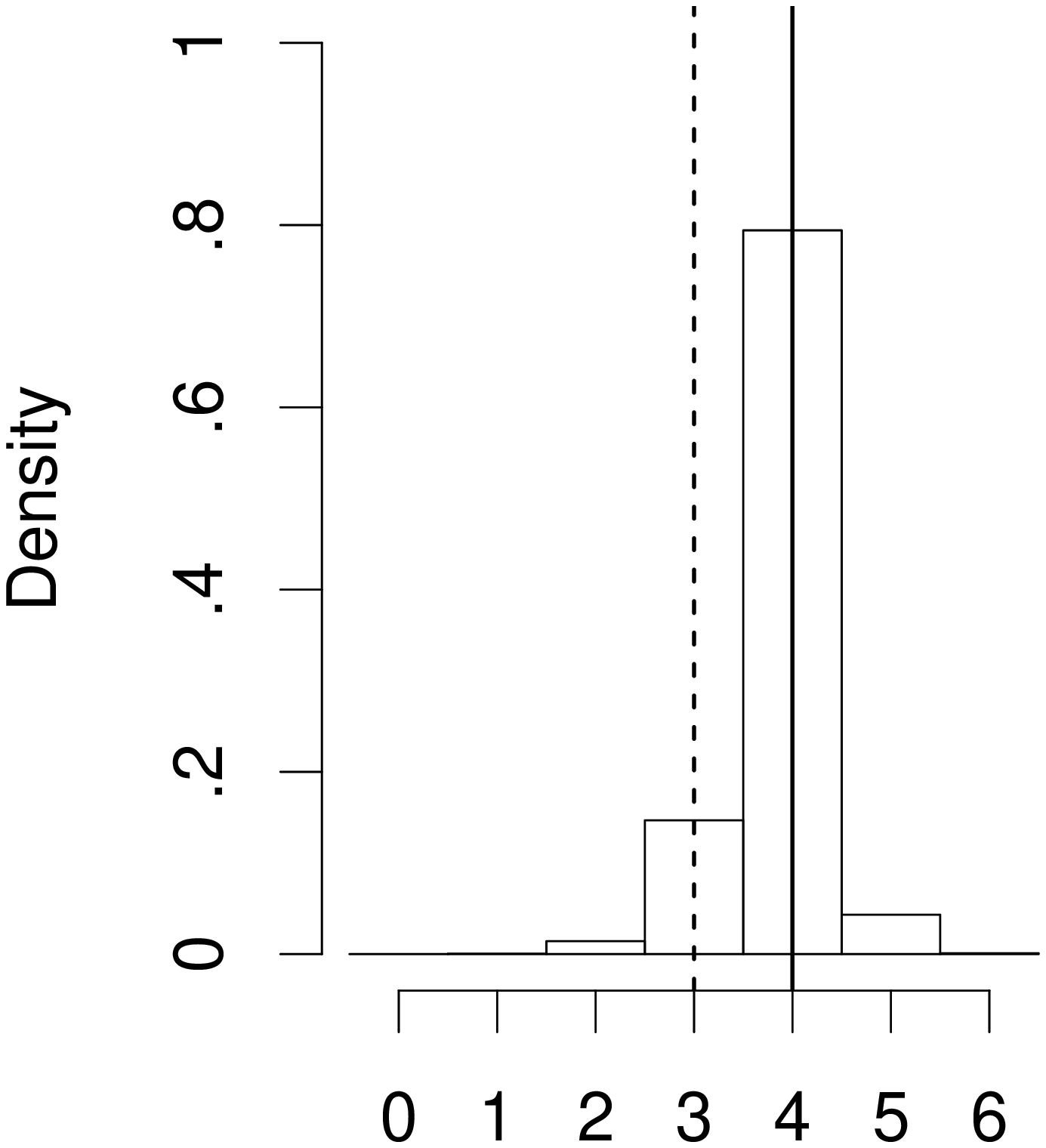}
\label{Zhist262}
}
\subfloat[$\bar{Z}_{65,2} = 8, \bar{\lambda} = 0.29$]{
\includegraphics[scale=.35]{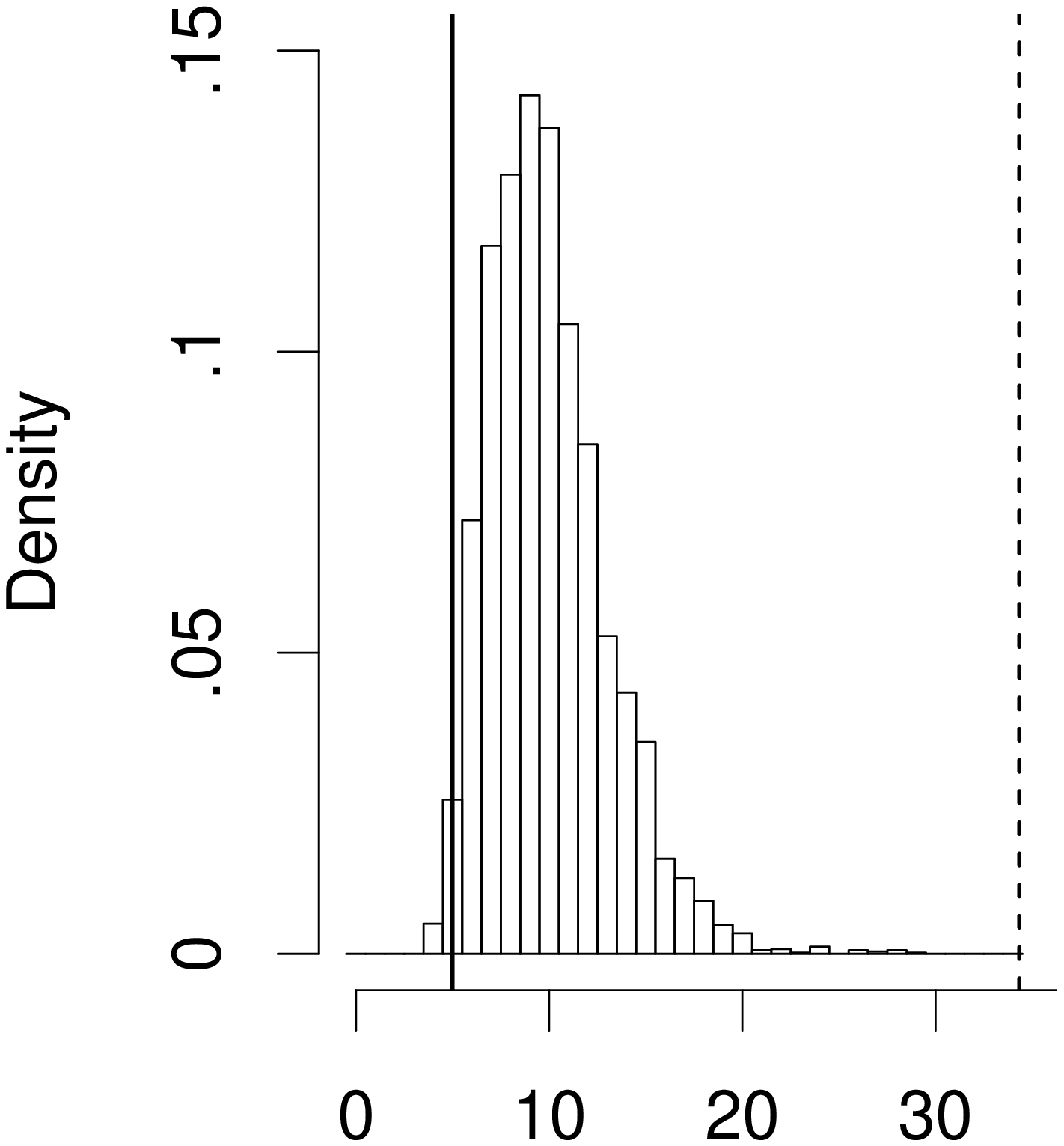}
\label{Zhist265}
}
\subfloat[$\bar{Z}_{65,3} = 67, \bar{\lambda} = 0.29$]{
\includegraphics[scale=.35]{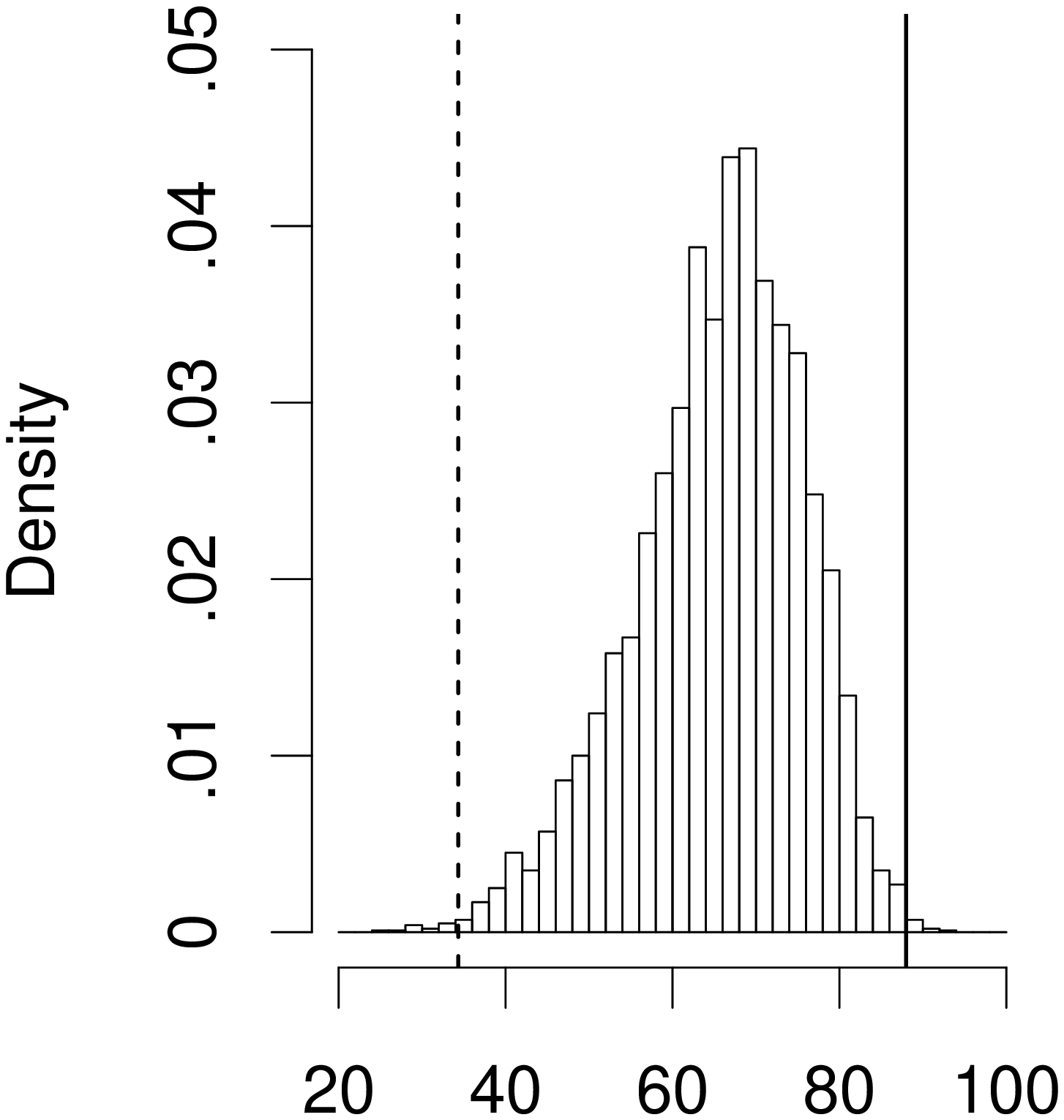}
\label{Zhist266}
}\\
\subfloat[$\bar{Z}_{65,3} = 86, \bar{\lambda} = 0.02$]{
\includegraphics[scale=.35]{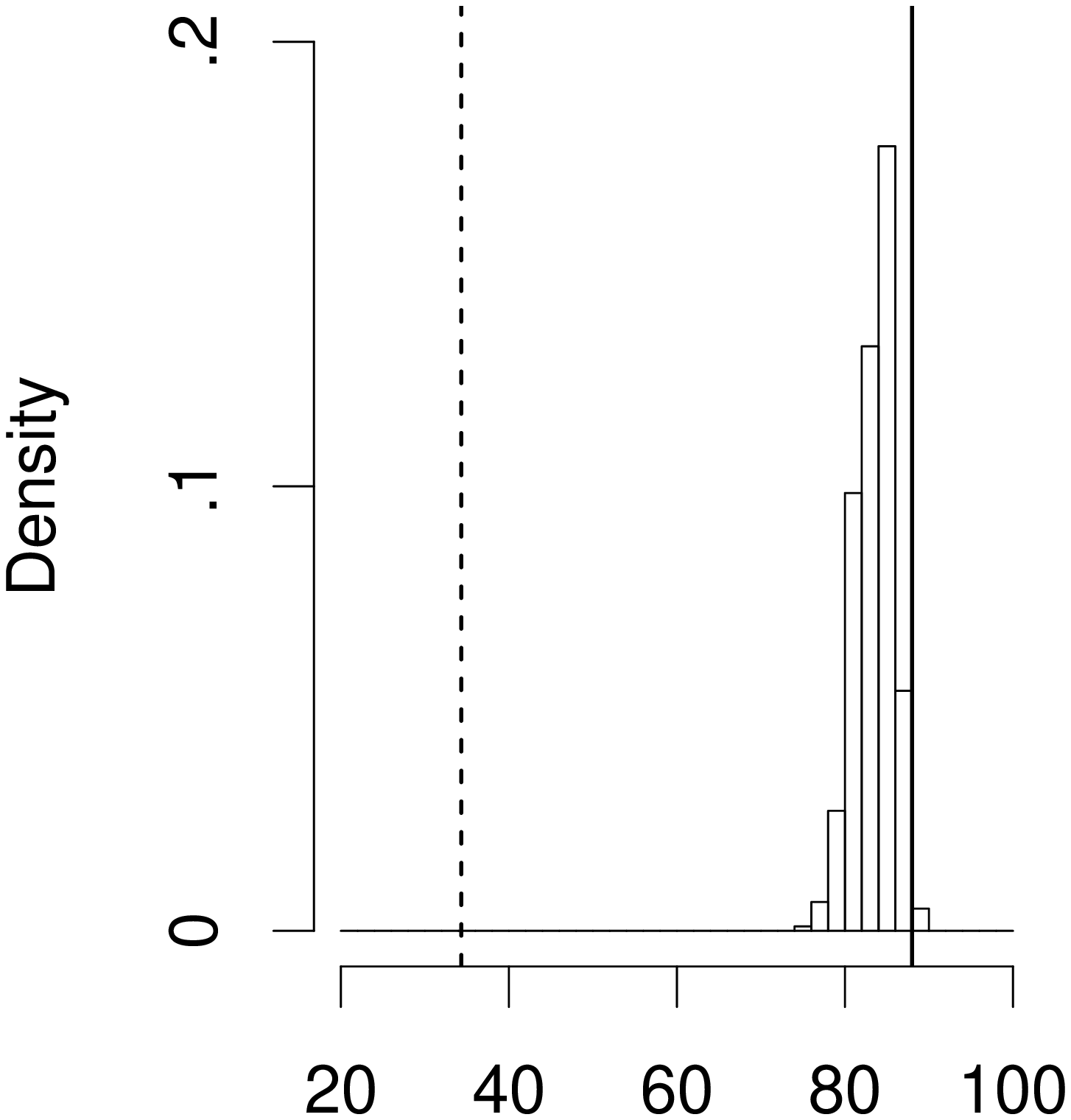}
\label{Zhist266fake}
}
\subfloat[$\bar{Z}_{92,2} = 13, \bar{\lambda} = 1.46$]{
\includegraphics[scale=.35]{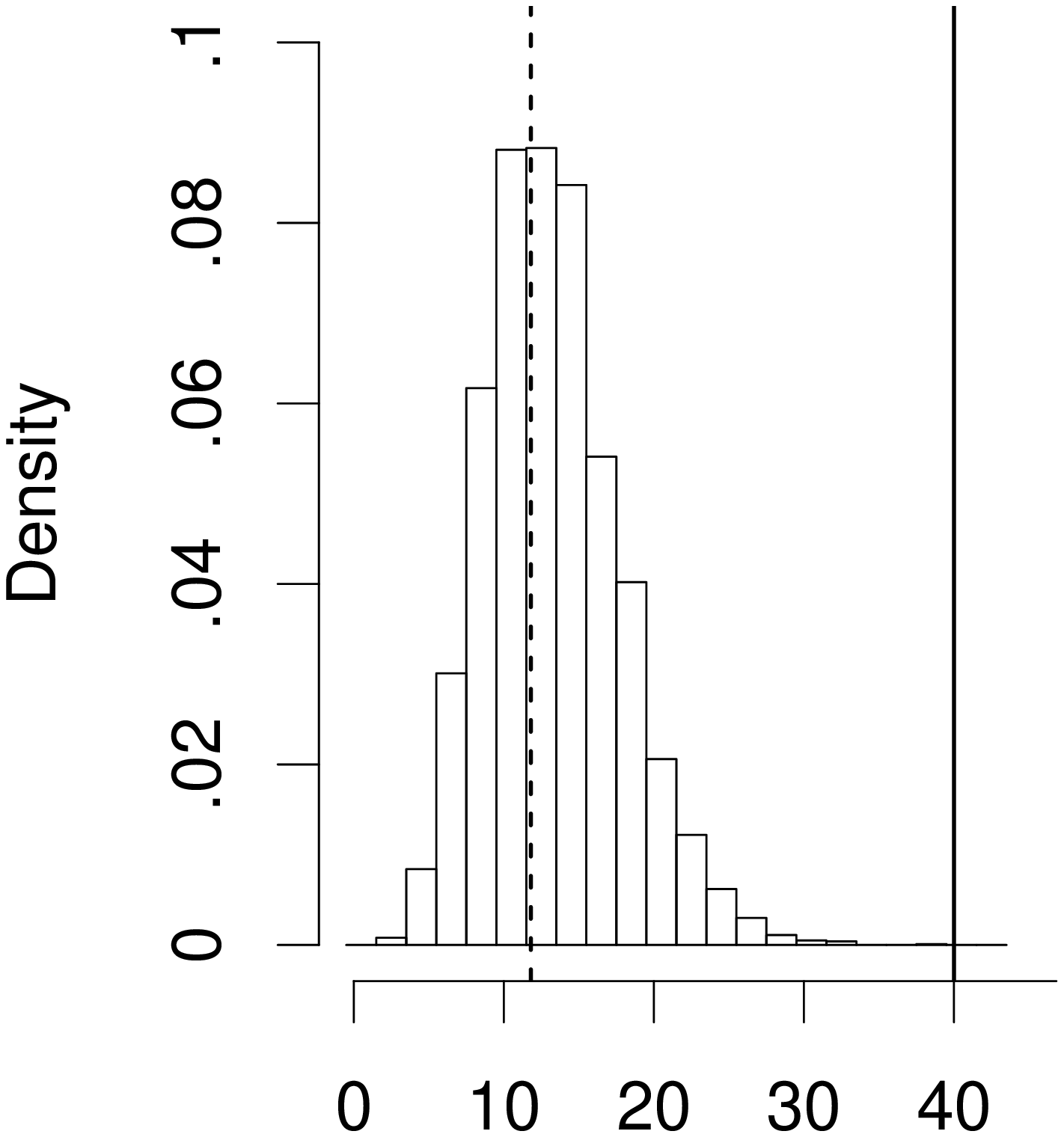}
\label{Zhist385}
}
\subfloat[$\bar{Z}_{88,4} = 68, \bar{\lambda} = 2.44$]{
\includegraphics[scale=.35]{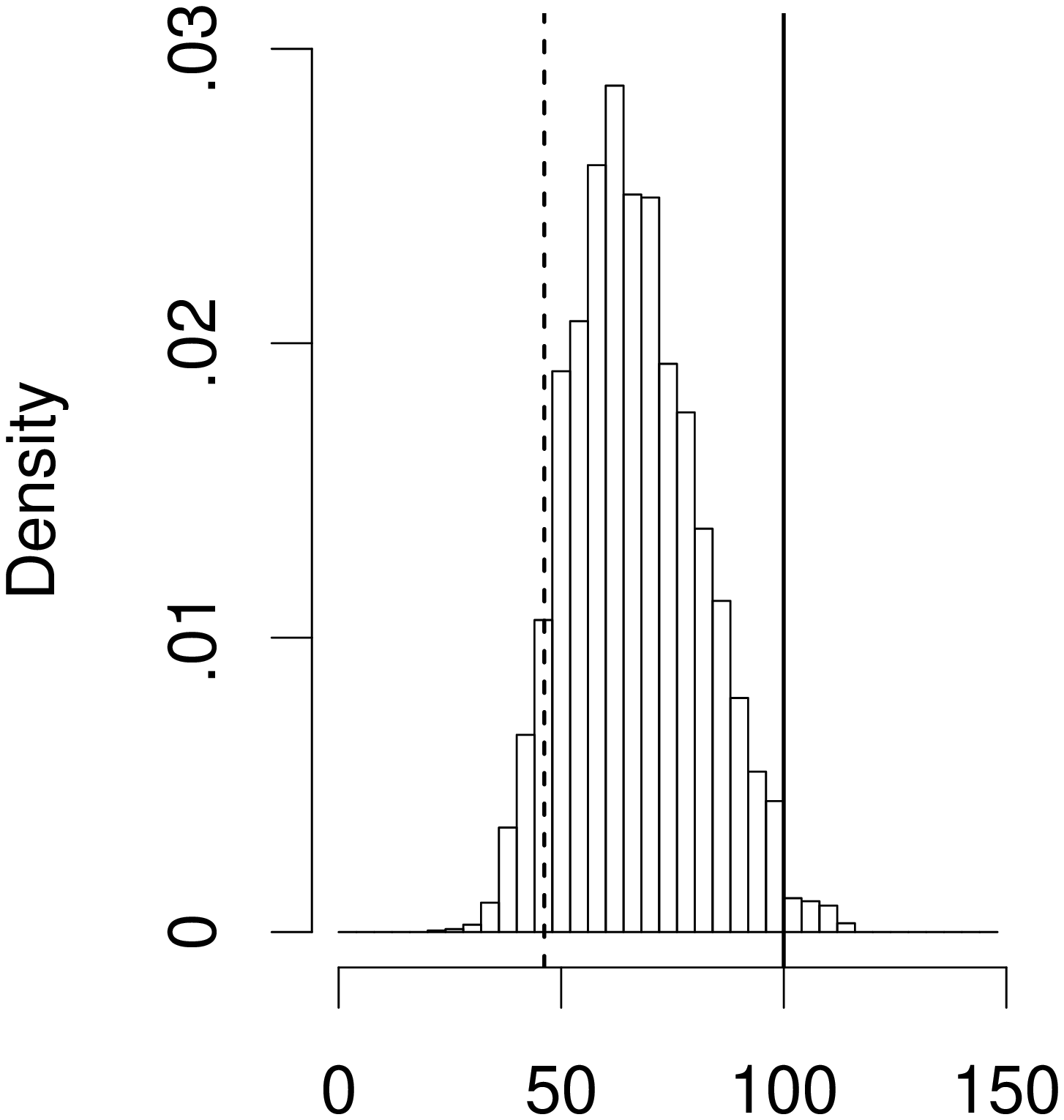}
\label{Zhist367}
}\\
\subfloat[$\bar{Z}_{95,4} = 14, \bar{\lambda} = 5.62$]{
\includegraphics[scale=.35]{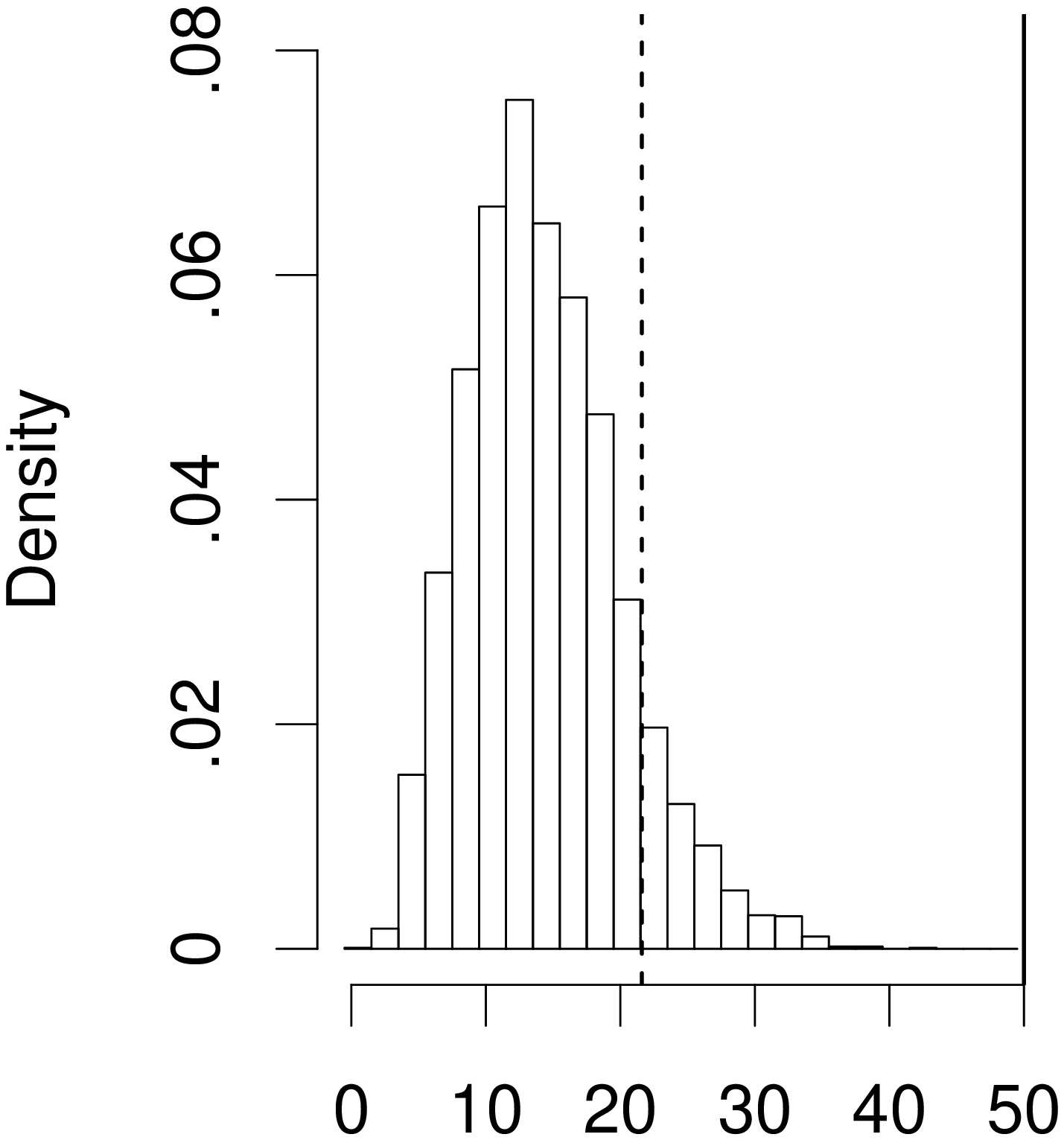}
\label{Zhist402}
}
\subfloat[$\bar{Z}_{116,1} = 20, \bar{\lambda} = 11.3$]{
\includegraphics[scale=.35]{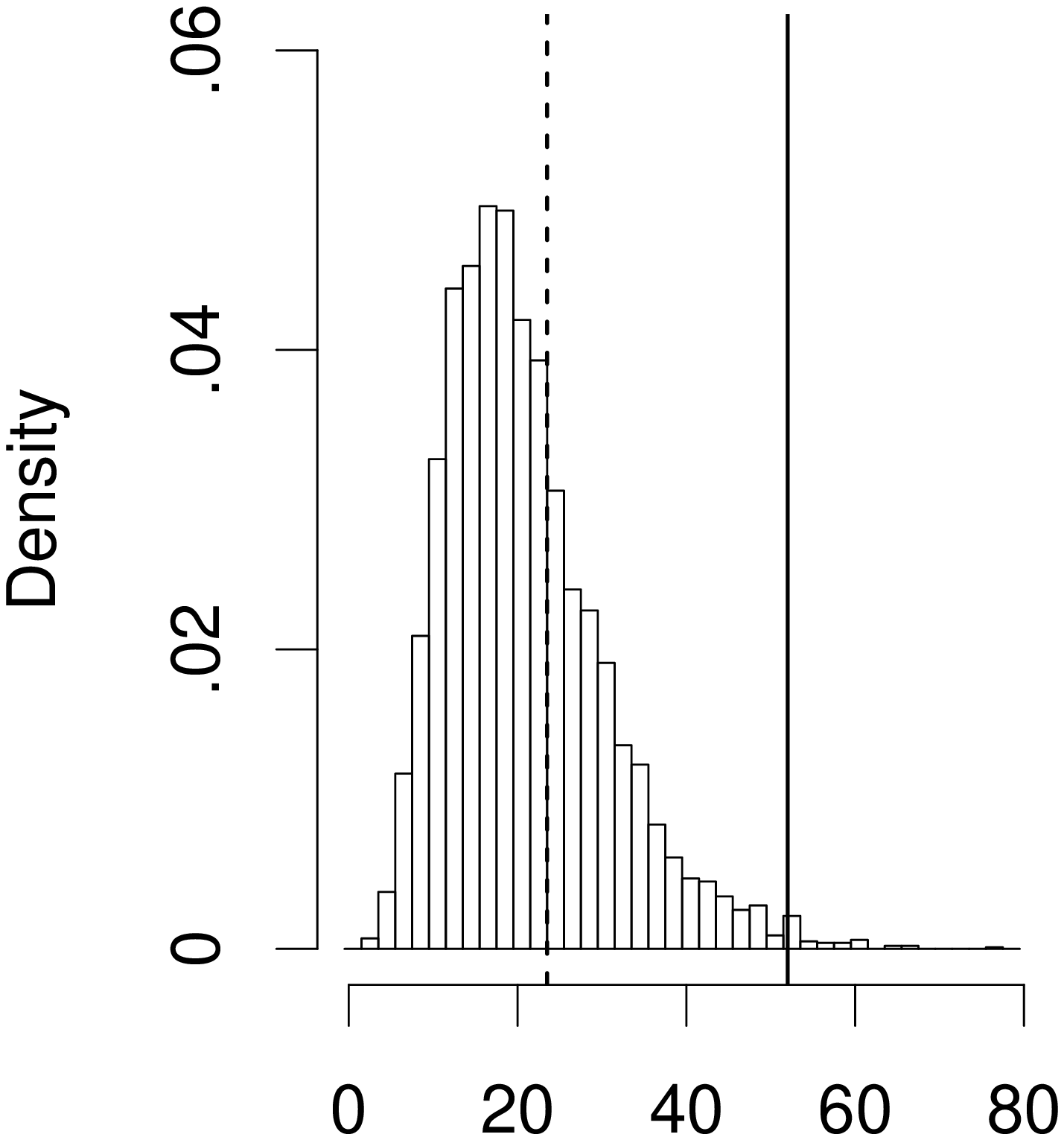}
\label{Zhist485}
}
\subfloat[$\bar{Z}_{44,1} = 18, \bar{\lambda} = 11.3$]{
\includegraphics[scale=.35]{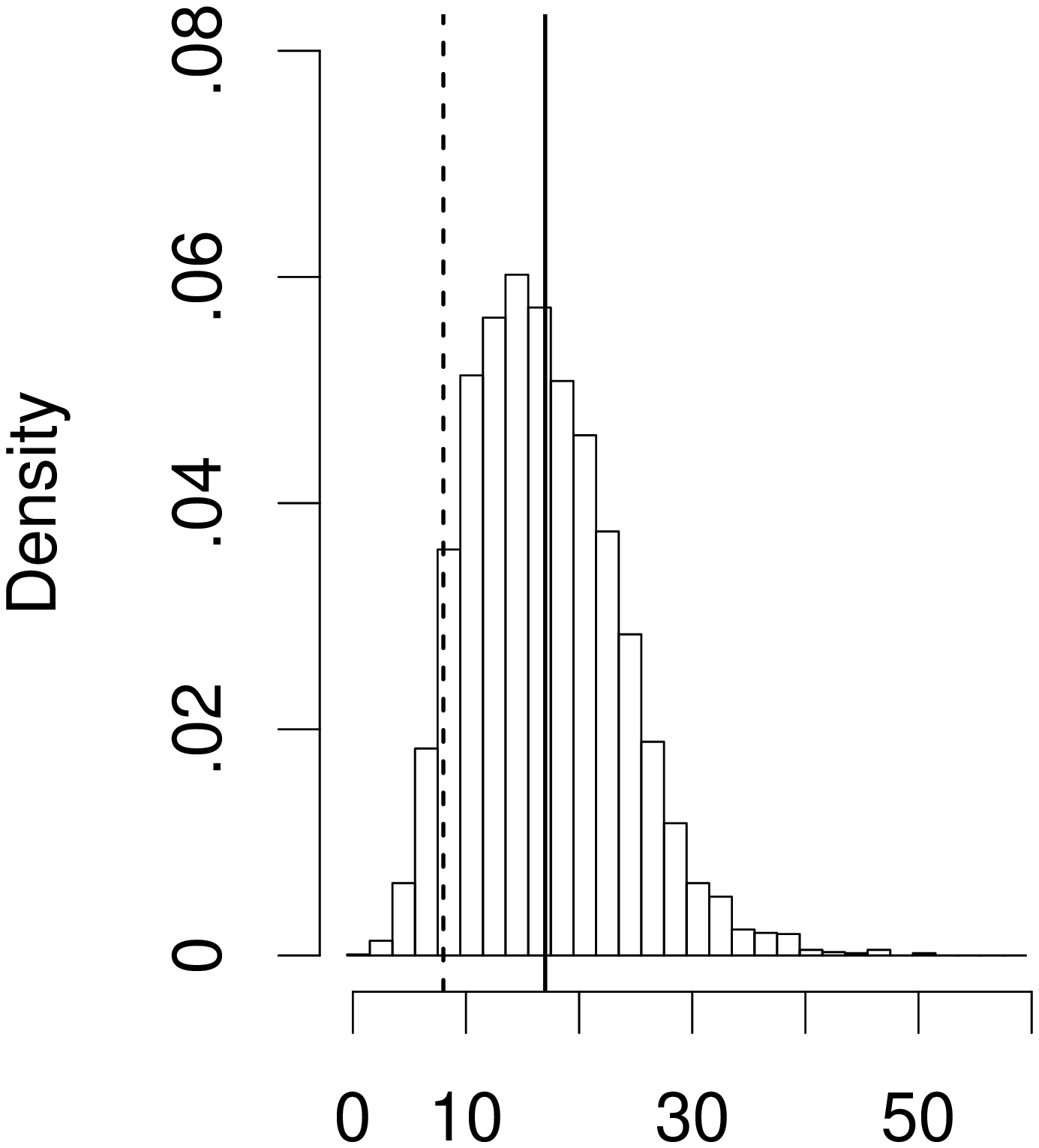}
\label{Zhist172}
}
\caption{\label{fig:ZpostDensity} Posterior density plots of selected true counts.
Vertical lines represent the reported count, and dashed lines represent the mean of observed counts for the corresponding subject.
$\bar{\lambda}$ is the estimated birth/death rate for the corresponding subject.}
\end{figure}

Figure~\ref{fig:ZpostDensity} presents posterior densities of 9 selected unobserved true counts $Z_{ij}$.  We chose these examples to illustrate various combinations of $Z_{ij}$ and $\lambda_{ij}$ values. In the figure, the solid vertical line in each plot identifies the reported count $Y_{ij}$ and the dashed vertical line reports the subject average $n_i^{-1}\sum_{j=1}^{n_i}Y_{ij}$ over time.
When the BD rate $\lambda_{ij}$ is close to zero, $Z_{ij}$ is close to the  $Y_{ij}$ as in Figure~\ref{fig:ZpostDensity}\subref{Zhist262} and \ref{fig:ZpostDensity}\subref{Zhist266fake}.
When $\lambda_{ij}$ is large, the variance of $Y_{ij}\mid Z_{ij}$ increases, and $Y_{ij}$ can be far from $Z_{ij}$ as in Figure~\ref{fig:ZpostDensity}\subref{Zhist485} and \ref{fig:ZpostDensity}\subref{Zhist172}.

Decomposing mean residual squared errors (MRSE) $m^{-1}\sum(Y_{ij}-\bar{\mu}_{ij})^{2}$, we have
\begin{align}
\label{eq:MRSE}
\frac{\sum(Y_{ij}-\bar{\mu}_{ij})^{2}}{m} & =
\frac{\sum(Y_{ij}-\bar{Z}_{ij})^{2}}{m} + \frac{\sum(\bar{Z}_{ij}-\bar{\mu}_{ij})^{2}}{m} \\ \nonumber
& + \frac{2\sum(Y_{ij}-\bar{Z}_{ij})(\bar{Z}_{ij}-\bar{\mu}_{ij})}{m},
\end{align}
where $\bar{Z}_{ij}$ and $\bar{\mu}_{ij}$ are the posterior means of $Z_{ij}$ and $\mu_{ij}$.
The first and the second terms on the right hand side can be interpreted as average measurement error and average Poisson sampling error, respectively.  However, the cross-product term is not zero. Table~{\ref{MSE}} presents decompositions of MRSE according to the ranges of $\bar{\lambda}_{ij}$, where $\bar{\lambda}_{ij}$ is the posterior means of $\lambda_{ij}$. In this manner, we see how the source of the variation differs depending on the BD rate.  When $\bar{\lambda}_{ij} < 0.05$, $\bar{Z}_{i,j} \approx Y_{ij}$ and so most of the variance in the data is from the Poisson random effects model.  When $\bar{\lambda}_{ij} \geq 1$, $\bar{Z}_{i,j} \approx \bar{\mu}_{ij}$ and so the BD process contributes the substantial portion of the variance.  If $\bar{\lambda}_{ij}$ is medium, $\bar{Z}_{i,j}$ lies in between $\bar{\mu}_{ij}$ and $Y_{ij}$ and the cross-product is the greatest contributor to the right side of \eqref{eq:MRSE}.

We learn from the BDPREM that IDU and MSM do not have more partners than non-IDU and heterosexual males or females, respectively.  The BDPREM tells us that subjects who engage in casual sex or sex trading have more partners than those who do not. As for the intervention effects, the BDPREM inference is of no overall intervention effect and no substantial difference between the two intervention modes.

\section{MODEL PERFORMANCE}
\label{simulation}

We conduct a simulation study to evaluate the performance of our proposed model and fitting procedure.  We employ the same mean model and predictor matrices from the CLEAR data set for the PREM and the BDPREM as those in the CLEAR data analysis. We generate random intercepts $\beta_{i}$ from a $N(0, .98)$ for $i=1, \ldots, 173$ where .98 is the rounded posterior mean of $D_\beta$ to 2 digit accuracy from the PREM in Table {\ref{estimates}}.  We take the posterior means of $\alpha_{1},\ldots,\alpha_{17}$ from the PREM in Table {\ref{estimates}} rounded to 2 digit accuracy as the true values in the simulation. We use the PS/PE prior for both models. True counts $Z_{ij}$ are then generated from a Poisson($\mu_{ij}$).  We use the values in the first column of Table {\ref{simulation psi}} for $\psi_{1},\ldots,\psi_{6}$ to calculate $\lambda_{ij}$.  Observed counts $Y_{ij}$ are generated from a BD process with initial state $Z_{ij}$ and BD rate $\lambda_{ij}$. In total, we generate 100 data sets.

Using the generated $Y_{ij}$'s, we fit our proposed BDPREM and PREM using the same algorithms as for the main data analysis but with 100,000 iterations following 10,000 burn-in iterations. Table {\ref{simulation1}} presents MSE, bias, variance of the posterior means averaged over the 100 analyses, and it presents the coverage proportion for the 95\% Bayesian credible intervals. The proposed BDPREM produces substantially lower variance and MSE for the regression coefficients $\alpha$ and for the random intercept variance $D_\beta$ than the PREM fit to the same data: the BDPREM MSE is less than 50\% of the MSE from the PREM model on average. Table {\ref{simulation psi}} shows that the mean of the 100 posterior means of $\bfpsi$ is close to the true $\bfpsi$. The bias of a parameter estimate is `significant' if the $t$ statistic calculated as bias divided by the square root of the (simulation variance divided by the number of simulations) is greater than 2 in absolute value. The $t$ statistic is mostly between .3 and .6 in absolute value for all parameters for the BDPREM and this means that the bias is explained by simulation variance. The two exceptions are for the PREM model: the intercept is biased low and $D_\beta$ is biased high.

We similarly fit the two models when the PREM is the true model using the generated $Z_{ij}$'s and the results are shown in Table {\ref{simulation2}}.  When the BDPREM is the true model, the PREM MSE averaged over the regression coefficients other than the intercept is 2.23 times greater than for the BDPREM.  On the other hand, when the PREM is the true model, the BDPREM and PREM have approximately the same MSE and thus, BDPREM does not lose any efficiency compared to PREM; estimates are not biased for either model.

\section{DISCUSSION}
\label{discussion}

We have presented a novel model for count data to explicitly account for reporting errors using a BD process. The proposed BDPREM is innovative because unlike most models such as random effects models, the BD process is defined on the same outcome space as the observables (i.e.\ integers), which eases interpretability as a benefit. The BD process variance is proportional to the hypothetical true count $Z_{ij}$, fitting with the finding that the accuracy of self-reported number of partners decreases with increasing number of partners (\citealt{jam04}).

The BD process can be generalized to have more complex properties and this represents an active area of research (\citealt{crawford2012transition, crawford2013estimation, doss2013fitting, crawford2014}). For example, in larger data sets with many repeated measures, one could consider replacing \eqref{eqloglinear} with a random effects model $\lambda_{ij} = \exp(\bfw_{ij}^{'} \bfpsi + \epsilon_{i})$ with $\epsilon_i \sim N(0,D_\epsilon)$ where $\epsilon_{i}$ is a subject-specific random intercept. The random effect $\epsilon_{i}$ allows an individual's BD rate to deviate from the population rate. Conveniently, the random effect also allows two subjects with equal covariates $\bfw_{ij}$ to have different $\lambda_{ij}$, and the random effect induces correlation among the BD rates within subject. The variance parameter $D_{\epsilon}$ can be modeled as having an inverse-Gamma distribution with shape parameter $a_{\epsilon}$ and scale parameter $b_{\epsilon}$: $D_{\epsilon} \sim IG(a_{\epsilon},b_{\epsilon})$. In our analysis, we did not have a subject random effect for the BD process, but we include the random effect in the posterior in Appendix B and in the computational algorithm presented in Appendix \ref{sampling algorithms} for generality.

Our story in this paper has been that the $Z_{ij}$ are \textit{unobserved true counts} while the $Y_{ij}$ are the reported counts. In practice, we do believe that people mis-report their numbers of sex partners, however, we are less sanguine about whether the unobserved true counts are actually modeled by the PREM. The truth is likely that true numbers of sex partners are naturally over-dispersed particularly in high risk populations, and that mis-reporting increases the over-dispersion. Thus even if we had the unobserved true counts, we would still need and prefer the BDPREM model over the PREM model. In this situation of mixed mis-reporting and natural over-dispersion, attribution of covariate effects in the BD rate model need to be taken with care. In the case of the CLEAR data and the results reported in table \ref{birth rate result}, we feel that the effects of CASUAL and TRADE likely reflect both mis-reporting and natural over-dispersion. On the other hand, the telephone treatment effect that shows people in the telephone intervention group are prone to significantly higher mis-reporting compared to the control group, the in-person intervention and baseline. We suspect this effect is mostly mis-reporting, perhaps related to the relatively alienating effect of intervention being delivered only through a cell-phone.

Finally, in terms of the CLEAR data, the BDPREM is important because it provides a much better fit to the data and we conclude that there are no intervention effects on the particular outcome while the poor-fitting PREM concludes that the interventions are effective.

\appendix
\begin{center}
\textbf{Appendix}
\end{center}
\makeatletter
 \renewcommand{\@seccntformat}[1]{APPENDIX~{\csname the#1\endcsname}.\hspace*{1em}}
 \makeatother

\section{Birth-Death Process Transition Probabilities}
\label{BD prob}

As many applied statisticians are unfamiliar with birth-death (BD) processes, we briefly review a derivation of the BD transition probabilities $p(S_{1}=y \mid S_{0}=z)$ for the restricted process exploited in this paper.  The probability generating function $G(s)$ of a random variable $S_{1}$ taking on non-negative integer values $y = 0,1,\ldots$ is defined as
\begin{align}
G(s) & \equiv E(s^{S_{1}} \mid S_{0} = z) \nonumber \\
& = \sum_{y=0}^{\infty} \mbox{Pr}(S_{1}=y \mid S_{0} = z)s^{y} . \label{eq:generating ft def}
\end{align}
For our model with equal birth and death rates, one can solve for $G(s)$ as the solution to a partial differential equation arising from the Chapman-Kolmogorov equation characterizing the process; interested readers should consult introductory texts in probability, such as \citet*{nor64} and \citet*{kar75}.
Our probability generating function becomes
\begin{equation}
G(s) = \left\{\frac{1-(\lambda-1)(s-1)}{1-\lambda(s-1)}\right\}^{z} \label{eq:generating ft}
\end{equation}
(\citealt*{nor64}).  Letting $\upsilon = \lambda/(1+\lambda)$ and expanding \eqref{eq:generating ft} in powers of $s^y$ yields the coefficients of $s^y$
\begin{equation}
\mbox{Pr}(S_{1}=y \mid 1, \lambda) = \left\{
\begin{aligned}
& (1 - \upsilon)^{2}\upsilon^{y-1}, \quad y \geq 1 \\
& \upsilon, \quad  y=0,
\end{aligned} \right.
\end{equation}
when $z=1$ (\citealt*{nor64}).
For $z\geq1$, expanding \eqref{eq:generating ft} using a Taylor series provides the more general solution
\begin{equation}
\mbox{Pr}(y \mid z, \lambda)  = \left\{
\begin{aligned}
& \displaystyle \sum_{j=1}^{\min(y,z)}\left(\begin{array}{c} z \\ j \end{array} \right) \left(\begin{array}{c} y-1 \\ j-1 \end{array} \right)\upsilon^{z+y-2j}(1-\upsilon)^{2j}, \quad y \geq 1  \\
& \upsilon^{z}, \quad y=0. \label{eq:bdprob}
\end{aligned} \right.
\end{equation}
Thus we see that the distribution of $y|z$ is a finite mixture of negative binomials.
When $z=0$, $S_1$ is 0 with probability 1.

\section{Posterior distribution}
\label{PosteriorDisti}

In our posterior formula and posterior sampling algorithm, we include the extension mentioned in the discussion that models the Birth-Death rate parameter $\lambda_{ij}$ with both fixed and random effects. The posterior distribution of the birth-death Poisson random effects model is
\begin{eqnarray}
\lefteqn{p(\bfZ,\bfalpha,\bfbeta,\bflambda,\bfpsi,\bfxi,\mathbf{D}_{\beta},D_{\epsilon} \mid \bfY,\bfX, \bfW)}   \nonumber \\
& & \propto p(\bfY \mid \bfZ,\bflambda)p(\bfZ \mid \bfalpha,\bfbeta,\bfX)p(\bfbeta \mid \mathbf{D}_{\beta})p(\bflambda \mid \bfY,\bfZ,\bfW,\bfpsi,\bfxi)p(\bfxi \mid D_{\epsilon}) \nonumber \\
& & \; \times \pi(\bfalpha)\pi(\bfpsi)\pi(\bfD_{\beta})\pi(D_{\epsilon}) \nonumber \\
& & \propto \displaystyle \prod_{i=1}^{n} \prod_{j=1}^{n_{i}}\left\{p(Y_{ij} \mid Z_{ij}, \lambda_{ij}) \frac{\exp\{Z_{ij}(\mathbf{x}_{ij}^{'}\bfalpha + \bfh_{ij}^{'}\bfbeta_{i})-\exp(\mathbf{x}_{ij}^{'} \bfalpha + \bfh_{ij}^{'}\bfbeta_{i})\}}{Z_{ij}!}\right\} \nonumber\\
& & \; \times \left|\bfD_{\beta}\right|^{-n/2} \exp \left(-\frac{1}{2}\sum_{i=1}^{n}\bfbeta_{i}^{'}\bfD_{\beta}^{-1}\bfbeta_{i}\right)
\times D_{\epsilon}^{-n/2} \exp\left(-\frac{1}{2D_{\epsilon}}\sum_{i=1}^{n}\epsilon_i^2\right) \nonumber \\
& & \; \times \left|\bfSigma_{\alpha}\right|^{-1/2}\exp\left(-\frac{1}{2} (\bfalpha - \bfalpha_{0})^{'}\bfSigma_{\alpha}^{-1}(\bfalpha - \bfalpha_{0})\right) \nonumber \\
& & \; \times \left|\bfSigma_{\psi}\right|^{-1/2}\exp\left(-\frac{1}{2} (\bfpsi - \bfpsi_{0})^{'}\bfSigma_{\psi}^{-1}(\bfpsi - \bfpsi_{0})\right) \nonumber \\
& & \; \times \left|\bfD_{\beta}\right|^{-(m_{\beta}+r+1)/2}\exp\left(-\mbox{tr}(\bfOmega_{\beta} \bfD_{\beta}^{-1})/2\right)
\times D_{\epsilon}^{-(a_{\epsilon}+1)}\exp^{-b_{\epsilon}/D_{\epsilon}}.
\end{eqnarray}

\section{Sampling Algorithms}
\label{sampling algorithms}

We use a Metropolis random walk algorithm (\citealt{met53}) for sampling from the posterior distributions of $\bfbeta$, $\bfalpha$, and $\bfpsi$ with a multivariate normal proposal with a diagonal covariance matrix.  For the $Z_{ij}$, we use a Metropolis-Hastings algorithm because the proposal distribution is not symmetric when the probability of moving from one point to the other point is not the same as the probability of the reverse movement due to boundary effects.  The variance or covariance matrix of the proposal distribution is multiplied by the scale parameter updated at each iteration using the auto-optimization algorithm.  We pick target acceptance probabilities of 0.2 to 0.4 as suggested in \citeauthor{gel04} (\citeyear{gel04}, chap.11).

\subsection{Sampling $\beta_i$}

We use hierarchical centering (\citealt*{gel95}) to sample $\beta_i$.
Let $\bfx_{i}^{(F)}$ and $\bfx_{i}^{(V)}$ denote time-fixed and time-varying covariate matrix for subject $i$.
An $n_{i} \times p$ covariate matrix $\bfx_{i}$ is partitioned as $(\bfx_{i}^{(F)},\bfx_{i}^{(V)})$ for all subjects $i$, separating time-fixed and time-varying covariates. Similarly $\bfalpha$ is partitioned as $(\bfalpha^{(F)}, \bfalpha^{(V)})$.  While the model is unchanged, $\beta_{i}$ is transformed to
\begin{align}
\eta_{i} = \beta_{i} + \bfx_{i1}^{(F)}\bfalpha^{(F)}, \label{eq:sampling beta}
\end{align}
where $\bfx_{i1}^{(F)}$ is the first row vector of $\bfx_{i}^{(F)}$.  Instead of sampling $\beta_{i}$ and $\bfalpha$, we sample $\eta_{i}$ and $\bfalpha$, and $\beta_{i}$ is obtained through \eqref{eq:sampling beta}.
The log conditional distribution of $\beta_{i}$ is
\begin{eqnarray}
\lefteqn{\log p(\beta_{i} \mid \bfZ, \bfalpha, D_\beta)} \nonumber \\
& = c + \displaystyle \sum_{j=1}^{n_{i}}\left[Z_{ij}(\bfx_{ij}^{'}\bfalpha + \beta_{i})-\exp(\bfx_{ij}^{'} \bfalpha + \beta_{i})\right] - \frac{1}{2D_\beta}\beta_{i}^{2}, \label{eq:logpost of beta}
\end{eqnarray}
where $c$ represents a fixed constant of proportionality that will vary with the equation.  Replacing $\bfx_{ij}^{'}\bfalpha + \beta_{i}$ with $\eta_{i}+\bfx_{ij}^{(V)}\bfalpha^{(V)}$ and $\beta_{i}$ with $\left(\eta_{i}-\bfx_{i1}^{(F)}\bfalpha^{(F)}\right)$ in  \eqref{eq:logpost of beta},
the log conditional distribution of $\eta_{i}$ is
\begin{align}
\lefteqn{\log p(\eta_{i} \mid \bfZ, \bfalpha^{(F)}, \bfalpha^{(V)}, D_\beta)} \nonumber \\
& = c + \displaystyle \sum_{j=1}^{n_{i}}\left(Z_{ij}\eta_{i} - \exp(\eta_{i}+\bfx_{ij}^{(V)}\bfalpha^{(V)})\right) -
\frac{1}{2D_\beta}\left(\eta_{i} - \bfx_{i1}^{(F)}\bfalpha^{(F)}\right)^{2}. \label{eq:logpost of eta}
\end{align}

\subsection{Sampling of $\bfalpha^{(V)}$}

The log conditional distribution of $\bfalpha^{(V)}$ is
\begin{eqnarray}
\lefteqn{\log p(\bfalpha^{(V)} \mid \bfZ,\bfeta,D_\beta)} \nonumber \\
& \propto \sum_{i,j} \left(Z_{ij}(\bfx_{ij}^{(V)}\bfalpha^{(V)}) - \exp(\eta_{i}+\bfx_{ij}^{(V)}\bfalpha^{(V)})\right) \nonumber \\
& - \frac{1}{2}(\bfalpha^{(V)} - \bfalpha_{0}^{(V)})^{'}\bfSigma_{\bfalpha^{(V)}}^{-1}(\bfalpha^{(V)} - \bfalpha_{0}^{(V)}),
\end{eqnarray}
where $\bfeta = (\eta_{1}, \ldots, \eta_{n})^{'}$ is an $n \times 1$ vector.

\subsection{Sampling of $\bfalpha^{(F)}$}

Taking advantage of hierarchical centering, the conditional posterior of $\bfalpha^{(F)}$ given $\bfeta$ and $D_\beta$ is a multivariate normal distribution
\begin{align}
\label{eq:sampling alphaF}
\bfalpha^{(F)} \mid \bfeta, D_\beta \sim \mbox{N}\left((\bfSigma_{\bfalpha^{(F)}}^{-1} +  D_\beta^{-1}{\bfX^{(F)}}^{\prime}\bfX^{(F)})^{-1}(D_\beta^{-1}{\bfX^{(F)}}^{\prime}\bfeta + \bfSigma_{\bfalpha^{(F)}}^{-1}\bfalpha^{(F)}_0), \right. \nonumber\\
\left.(\bfSigma_{\bfalpha^{(F)}}^{-1} + D_\beta^{-1}{\bfX^{(F)}}^{\prime}\bfX^{(F)})^{-1}\right),
\end{align}
where $\bfX^{(F)} = ({\bfx_{11}^{(F)}}^{\prime}, \ldots, {\bfx_{n1}^{(F)}}^{\prime})^{\prime}$ and $\bfx_{i1}^{(F)}$ is the first row of $\bfx_{i}^{(F)}$.

\subsection{Sampling $D_\beta^{-1}$}

For simplicity, define $K = D_\beta^{-1}$.  $K$ has conditional density
\begin{align}
K \mid \bfbeta \sim \mbox{Gamma}\left(\frac{n}{2}+ a,\frac{1}{2}\bfbeta^{\prime}\bfbeta+ b \right),
\end{align}
where $\mbox{Gamma}(\cdot,\cdot)$ denotes a gamma distribution and $\bfbeta$ is an $n \times 1$ vector.  The pdf of a gamma distribution for $x > 0$ is $f(x\mid k,\theta) = x^{k-1}\theta^{k}\exp(-\theta x)/\Gamma(k)$.

\subsection{Sampling $Z_{ij}$}

We update one $Z_{ij}$ at a time.  The log posterior density is
\begin{align}
\lefteqn{\log p(Z_{ij} \mid \bfalpha,\bfeta,\bflambda,\bfY)} \nonumber \\
& = c + \log \left[\mbox{Pr}(Y_{ij} \mid Z_{ij}, \lambda_{ij})\right] + Z_{ij}(\eta_{i} + \bfx_{ij}^{(V)}\bfalpha^{(V)}) - \log (Z_{ij}!),
\end{align}
where $\bflambda = (\lambda_{11},\ldots,\lambda_{nn_{n}})^{'}$ is an $N \times 1$ vector and $\mbox{Pr}(Y_{ij} \mid Z_{ij}, \lambda_{ij})$ is given in \eqref{eq:bdprob}. The first term arises from the BD process and the last two terms are from the PREM.  The scalar $\log Z_{ij}!$ for large $Z_{ij}$ is calculated as $\log [\Gamma(Z_{ij} + 1)]$, where $\Gamma(n)=\int_0^\infty x^{n-1}e^{-x} dx$ is the Gamma function.
We sample $Z_{ij}$ through a Metropolis-Hastings algorithm. Since $Z_{ij}$ is a non-negative integer, the transition distribution should be on the integers.  Define $Z_{ij}^{(l)}$ as the $l$th sample for subject $i$ at time $t_{ij}$.  The $(l+1)$st sample proposal $Z_{ij}^{(l+1)*}$ is sampled differently depending on the values of $Z_{ij}^{(l)}$ and $Y_{ij}$:
\begin{enumerate}
\renewcommand{\theenumi}{\roman{enumi}}
  \item If $Z_{ij}^{(l)} = 0$, then the jump is either 0 or 1 with each probability of 0.5 giving $Z_{ij}^{(l+1)} = 0 \mbox{ or } 1$.
  \item If $Z_{ij}^{(l)} = 1$ and $Y_{ij} > 0$, then the jump is either 0 or 1 each with probability 0.5 giving $Z_{ij}^{(l+1)} = 1 \mbox{ or } 2$.  When $Y_{ij} > 0$, the sample $Z_{ij}^{(l+1)} = 0$ is not allowed because neither births nor deaths can occur from a zero state, i.e.\ $\mbox{Pr}(Y_{ij} > 0 \mid Z_{ij} = 0) = 0$.
  \item If $Z_{ij}^{(l)} = 1$ and $Y_{ij} = 0$, then the transition distribution of $Z_{ij}^{(l+1)*} - Z_{ij}^{(l)}$is a discrete uniform distribution with support $\{-1,0,1\}$.
  \item If $Z_{ij}^{(l)} > 1$, then we allow a more flexible range for the jump.  We allow the support of the discrete uniform distribution to be on the integers between $-\left\lceil Z_{ij}/2 \right\rceil$ and $\left\lceil Z_{ij}/2 \right\rceil$, where $\lceil x \rceil$ is the ceiling function, defined as the smallest integer greater than $x$.
\end{enumerate}
Thus, the proposal density $g(u \mid v)$ for the transition from $v$ to $u$ is
\[g(u \mid v) = \left\{
\begin{array}{l l l l l}
\frac{1}{2} & \quad \mbox{if $v = 0$} \\
\frac{1}{2} & \quad \mbox{if $v = 1$ and $u = 1$ or 2 and $Y_{ij} > 0$} \\
\frac{1}{3} & \quad \mbox{if $v = 1$ and $u = 0,1,$ or 2 and $Y_{ij} = 0$}\\
\frac{1}{2\left\lceil \frac{v}{2} \right\rceil + 1} & \quad \mbox{if $v > 1$ and $v-\left\lceil\frac{v}{2}\right\rceil \leq u \leq v+\left\lceil\frac{v}{2}\right\rceil$}\\
0 & \quad \mbox{otherwise}.\end{array} \right. \]
Following this algorithm, sample a candidate $Z_{ij}^{(l+1)*}$, and compute the Metropolis-Hastings ratio $R(Z_{ij}^{(l)}, Z_{ij}^{(l+1)*})$
\begin{align}
R(Z_{ij}^{(l)}, Z_{ij}^{(l+1)*}) = \min\left(1,\frac{p(Z_{ij}^{(l+1)*} \mid \bfalpha,\bfeta,\bflambda,\bfY)}{p(Z_{ij}^{(l)} \mid \bfalpha,\bfeta,\bflambda,\bfY)} \times \frac{g(Z_{ij}^{(l)} \mid Z_{ij}^{(l+1)*})}{g(Z_{ij}^{(l+1)*} \mid Z_{ij}^{(l)})}\right).
\end{align}
Generate a random number $U \sim \mbox{Uniform}[0,1]$, and accept $Z_{ij}^{(l+1)*}$ if $U < R(Z_{ij}^{(l)}, Z_{ij}^{(l+1)*})$ and reject otherwise.  The proposal density is not symmetric so that $g(Z_{ij}^{(l)} \mid Z_{ij}^{(l+1)*}) \neq g(Z_{ij}^{(l+1)*} \mid Z_{ij}^{(l)})$ for some pairs of $Z_{ij}^{(l)}$ and $Z_{ij}^{(l+1)*}$.

\subsection{Sampling $\bfpsi$}

The BDPREM (\ref{eqloglinear}) as expanded in the discussion section, second paragraph, contains random effects parameters.  Let $\bfpsi = (\psi_{1}, \ldots, \psi_{q})^{'}$ is a $q \times 1$ vector of regression coefficients.  The log conditional posterior of $\bfpsi$ given $\bfY$, $\bfZ$ and $\bflambda$ is
\begin{align}
\lefteqn{\log p(\psi \mid \bfY, \bfZ)} \nonumber \\
& = c + \sum_{i,j} \log \left(\mbox{Pr}(Y_{ij} \mid Z_{ij}, \lambda_{ij})\right) - \frac{1}{2}(\bfpsi - \bfpsi_{0})^{'}\bfSigma_{\psi}^{-1}(\bfpsi - \bfpsi_{0}).
\end{align}

\subsection{Sampling $\epsilon_i$}

Subject-specific random intercept $\epsilon_i$ for the BD process is sampled through the log conditional posterior
\begin{align}
\lefteqn{\log p(\epsilon_i \mid \bfY, \bfZ, \bfpsi, D_\epsilon)} \nonumber \\
& = c + \sum_j [\log p(Y_{ij} \mid Z_{ij}, \lambda_{ij})] - (2D_\epsilon)^{-1}\epsilon_i^2,
\end{align}
where $\lambda_{ij} = \exp(\bfw_{ij}^{'}) \bfpsi + \epsilon_i$ rather than as given in (\ref{eqloglinear}).

\subsection{Sampling $D_\epsilon^{-1}$}

For simplicity, define $H = D_\epsilon^{-1}$.  $H$ has conditional density
\begin{align}
H \mid \bfepsilon \sim \mbox{Gamma}\left(\frac{n}{2}+ a_\epsilon,\frac{1}{2}\bfepsilon^{\prime}\bfepsilon+ b_\epsilon \right),
\end{align}
where $\bfepsilon = (\epsilon_1,\ldots,\epsilon_n)^{'}$.

\bibliographystyle{asa}
\bibliography{reference}
\clearpage

\begin{landscape}
\begin{table}
\small
\caption{Prior and posterior parameter estimates for the Poisson random effects model (PREM) component of the BDPREM and the PREM models.  Columns 2-7 give posterior means and standard deviations (SDs) for the birth-death Poisson random effects model (BDPREM) with the previous study/pure elicitation (PS/PE), data augmentation (DA), and previous data set (DS) prior. Columns 8-9 give posterior means and SDs for the PREM with the PS/PE prior.
The last six columns are the PS/PE, DA, and DS prior means and SDs for the PREM component.  IDU is an indicator of injection drug use, MSM is an indicator of men who have sex with men. CASUAL and TRADE are indicators of engagement in casual sex and trading sex for money, drugs, food or housing in the past 3 months. Parameters labeled I*Month 3, 6, 9, and 15 and T*Month 3, 6, 9, and 15 are interactions between the in-person and telephone intervention group and the given follow-up month. Parameter $D_\beta$ is the variance of the random intercept.}

\begin{tabular}{lcccccccccccccccc}
\hline \hline
& \multicolumn{6}{c}{PREM with BD} & \multicolumn{1}{c}{} & \multicolumn{2}{c}{PREM} & \multicolumn{1}{c}{} & \multicolumn{6}{c}{Prior}\\
& \multicolumn{2}{c}{PS/PE} & \multicolumn{2}{c}{DA} & \multicolumn{2}{c}{DS} & \multicolumn{1}{c}{} & \multicolumn{2}{c}{PS/PE} & \multicolumn{1}{c}{} &
\multicolumn{2}{c}{PS/PE} & \multicolumn{2}{c}{DA} & \multicolumn{2}{c}{DS} \\
\cline{2-7} \cline{9-10} \cline{12-17}
Parameter & Mean & SD & Mean & SD  & Mean & SD & & Mean & SD & & Mean & SD & Mean & SD & Mean & SD \\ \hline
Intercept & -0.29 & 0.16 & -0.30 & 0.15 & -0.27 & 0.14 & & -0.42 & 0.17 & & 0 & 1.74  & -0.62 & 1.67 & -0.24 & 0.57\\
IDU & 0.16 & 0.19  & 0.18 & 0.18 & 0.14 & 0.17 & & 0.48 & 0.24 & & 0.78 & 1.33  & 1.05 & 2.10 & 0.21 & 0.77\\
MSM & 0.30 & 0.16  & 0.32 & 0.16 & 0.27 & 0.15 & & 0.60 & 0.20  & & 0.03 & 1.37  & 1.06 & 2.12 & 0.24 & 0.62\\
CASUAL & 1.57 & 0.10  & 1.57 & 0.10 & 1.58 & 0.09 & & 1.16 & 0.07 & & 1.25 & 1.10  & 1.90 & 2.00 & 1.23 & 0.49\\
TRADE & 1.13 & 0.13  & 1.15 & 0.13 & 1.13 & 0.13 & & 0.92 & 0.06 & & 1.2 & 1.53 &  2.63 & 1.98 & 0.54 & 0.90\\
Month 3 & -0.11 & 0.18  & -0.12 & 0.17 & -0.12 & 0.17 & & -0.32 & 0.11 & & 0 & 2 &  0.05 & 2.32 & -0.09 & 0.78\\
Month 6 & -0.28 & 0.18  & -0.31 & 0.18 & -0.30 & 0.18 & & -0.60 & 0.08 & & 0 & 2 &  0.01 & 2.31 & -0.43 & 0.85\\
Month 9 & -0.46 & 0.19  & -0.52 & 0.20 & -0.43 & 0.19 & & -0.87 & 0.11 & & 0 & 2 &  0 & 2.29 & -0.35 & 1.18\\
Month 15 & -0.32 & 0.18  & -0.36 & 0.20 & -0.31 & 0.19 & & -0.60 & 0.10 & & 0 & 2 & 0.08 & 2.31 & -0.27 & 1.12\\
I*Month 3 & -0.33 & 0.23  & -0.33 & 0.22 & -0.30 & 0.22 & & -0.65 & 0.16 & & 0 & 2 & -1.45 & 2.96 & 0.05 & 1.03\\
I*Month 6 & -0.11 & 0.24  & -0.10 & 0.25 & -0.09 & 0.24 & & -0.27 & 0.14 & & 0 & 2 & -1.32 & 2.89 & 0.23 & 1.14\\
I*Month 9 & 0.08 & 0.25  & 0.13 & 0.26 & 0.05 & 0.24 & & 0.35 & 0.17 & & 0 & 2 & -1.37 & 2.95 & 0.05 & 1.36\\
I*Month 15 & -0.05 & 0.24  & -0.04 & 0.26 & -0.08 & 0.25 & & -0.25 & 0.15 & & 0 & 2 & -1.43 & 2.99 & -0.06 & 1.49\\
T*Month 3 & 0.08 & 0.23  & 0.09 & 0.23 & 0.09 & 0.22 & & 0.82 & 0.13 & & 0 & 2 & -0.40 & 2.44 & 0.05 & 1.03\\
T*Month 6 & 0.26 & 0.24  & 0.28 & 0.24 & 0.28 & 0.24 & & 0.48 & 0.13 & & 0 & 2 & -0.33 & 2.41 & 0.23 & 1.14\\
T*Month 9 & 1.08 & 0.22  & 1.12 & 0.23 & 1.01 & 0.20 & & 1.66 & 0.13 & & 0 & 2 & -0.30 & 2.36 & 0.05 & 1.36\\
T*Month 15 & 0.07 & 0.25 & 0.11 & 0.27 & 0.06 & 0.26 & & 0.58 & 0.14 & & 0 & 2 & -0.45 & 2.45 & -0.06 & 1.49\\ \hline
$D_\beta$ & 0.43 & 0.07 & 0.45 & 0.08 & 0.39 & 0.07 & & 0.98 & 0.13  & & 1 & 1 & 1.02 & 1.05 & 0.27 & 0.27\\ \hline
\end{tabular}
\label{estimates}
\end{table}
\end{landscape}
\clearpage

\begin{landscape}
\begin{table}
\small
\centering
\caption{Birth-death component parameter estimates. The first six columns are posterior means and SDs for the BD model from the BDPREM using the PS/PE, DA, and DS prior, and the last six columns are prior means and SDs from the PS/PE, DA, and DS models. PB is post-baseline. I and T are indicators of being in the in-person intervention and telephone intervention.}
\begin{tabular}{lccccccccccccc}
\hline \hline
& \multicolumn{6}{c}{Posterior} & \multicolumn{1}{c}{} & \multicolumn{6}{c}{Prior} \\
& \multicolumn{2}{c}{PS/PE} & \multicolumn{2}{c}{DA} & \multicolumn{2}{c}{DS} & \multicolumn{1}{c}{} & \multicolumn{2}{c}{PS/PE} & \multicolumn{2}{c}{DA} & \multicolumn{2}{c}{DS}\\
\cline{2-7} \cline{9-14}
Parameter & Mean & SD & Mean & SD & Mean & SD & & Mean & SD & Mean & SD & Mean & SD\\ \hline
Intercept & -2.43 & 0.41  & -2.36 & 0.40 & -2.53 & 0.44 & &  0 & 2.24  & -1.01 & 1.52 & -2.10 & 1.34\\
PB & -0.68 & 0.22 & -0.73 & 0.22 & -0.72 & 0.22 & & 0 & 1.53  & 0.02 & 2.16 & -0.07 & 1.83\\
PB*I & -0.61 & 0.30  & -0.53 & 0.30 & -0.50 & 0.29 & & -0.69 & 1.18  & -0.25 & 2.23 & -0.54 & 1.57\\
PB*T & 0.51 & 0.23  & 0.58 & 0.23 & 0.56 & 0.24 & & -1.5 & 1.59 &  -0.02 & 2.15 & -0.54 & 1.57\\
CASUAL & 3.47 & 0.43  & 3.41 & 0.42 & 3.59 & 0.46 & & 1.85 & 1.06 &  1.28 & 1.93 & 1.81 & 1.26\\
TRADE & 1.36 & 0.18  & 1.38 & 0.18 & 1.35 & 0.18 & & 1.85 & 1.06  &  1.91 & 2.02 & 2.87& 1.20\\\hline
\end{tabular}
\label{birth rate result}
\end{table}
\end{landscape}
\clearpage

\begin{table}
\small
\caption{DA prior data for the PREM and BD components of the BDPREM. Columns $Y$ and $Z$ represent reported and true count.  The Intv column indicates the intervention group with `C',`I', and `T' representing  control, in-person, and telephone intervention, respectively.}
\begin{tabular}{cccccccc}
\hline
\multicolumn{8}{l}{DA prior for PREM} \\
\hline \hline
$Z$ & Intercept & IDU & MSM & CASUAL & TRADE & Intv & Time \\
\hline
1 & 1 & 0 & 0 & 0 & 0 & C & 0 \\
1 & 1 & 0 & 0 & 0 & 0 & C & 3 \\
1 & 1 & 0 & 0 & 0 & 0 & C & 6 \\
1 & 1 & 0 & 0 & 0 & 0 & C & 9 \\
1 & 1 & 0 & 0 & 0 & 0 & C & 15 \\
0.5 & 1 & 0 & 0 & 0 & 0 & I & 3 \\
0.5 & 1 & 0 & 0 & 0 & 0 & I & 6 \\
0.5 & 1 & 0 & 0 & 0 & 0 & I & 9 \\
0.5 & 1 & 0 & 0 & 0 & 0 & I & 15 \\
0.8 & 1 & 0 & 0 & 0 & 0 & T & 3 \\
0.8 & 1 & 0 & 0 & 0 & 0 & T & 6 \\
0.8 & 1 & 0 & 0 & 0 & 0 & T & 9 \\
0.8 & 1 & 0 & 0 & 0 & 0 & T & 15 \\
2 & 1 & 1 & 0 & 0 & 0 & C & 0 \\
2 & 1 & 0 & 1 & 0 & 0 & C & 0 \\
4 & 1 & 0 & 0 & 1 & 0 & C & 0 \\
8 & 1 & 0 & 0 & 0 & 1 & C & 0 \\
\hline
\multicolumn{8}{l}{DA prior for BD model} \\
\hline \hline
$Y$ & $Z$ & Intercept & PB & PB*T & PB*I & CASUAL & TRADE \\
\hline
5 & 2 & 1 & 1 & 1 & 0 & 0 & 0 \\
1 & 3 & 1 & 0 & 0 & 0 & 0 & 0 \\
1 & 1 & 1 & 1 & 0 & 1 & 0 & 0 \\
40 & 30 & 1 & 0 & 0 & 0 & 0 & 1 \\
7 & 10 & 1 & 1 & 0 & 0 & 0 & 0 \\
30 & 20 & 1 & 0 & 0 & 0 & 1 & 0 \\
\hline
\end{tabular}
\label{DA for PREM}
\end{table}
\clearpage

\begin{table}
\small
\caption{Decompositions of mean residual squared error (MRSE) according to the ranges of $\bar{\lambda}_{ij}$'s, where $\bar{\lambda}_{ij}$ is the posterior mean of $\lambda_{ij}$, $m$ is the number of $\bar{\lambda}_{ij}$'s falling in the given range.  The percentages are fractions of the $\mbox{MRSE}=\sum(Y_{ij}-\bar{\mu}_{ij})^{2}/m$, $\mbox{measurement error}=\sum(Y_{ij}-\bar{Z}_{ij})^{2}/m$, $\mbox{sampling error}=\sum(\bar{Z}_{ij}-\bar{\mu}_{ij})^{2}/m$, and $\mbox{cross product}=2\sum(Y_{ij}-\bar{Z}_{ij})(\bar{Z}_{ij}-\bar{\mu}_{ij})/m$, where $Y_{ij}$ is a reported count, $\bar{\mu}_{ij}$ is a mean number of partners estimated from the BDPREM, and $\bar{Z}_{ij}$ is a posterior mean of the latent true count $Z_{ij}$.}
\begin{tabular}{lccccc}
\hline \hline
&  $m$ & MRSE & measurement error & sampling error & cross product \\ \hline
$\bar{\lambda}_{ij} < 0.05$ & 211 & 0.32 & 0.001 (0.4\%) & 0.30 (94\%) & 0.02 (5.1\%) \\
$0.05 \leq \bar{\lambda}_{ij} < 1$ & 250 & 14.56 & 3.81 (26\%) & 4.07 (27\%) & 6.69 (46\%) \\
$\bar{\lambda}_{ij} \geq 1$ & 270 & 399 & 343.6 (86\%) & 3.74 (0.9\%) & 51.8 (13\%) \\ \hline
\end{tabular}
\label{MSE}
\end{table}
\clearpage

\begin{landscape}
\begin{table}
\small
\caption{Comparison of mean square error (MSE), bias, variance (Var), average over the posterior variances from the 100 simulated data sets (Avg Var), 95\% confidence coverage probability (CP) for the regression coefficients $\alpha$ of the PREM component and the variance $D_\beta$ of the random intercept from the BDPREM and the PREM, using data generated from BDPREM. I and T are indicators of being in the in-person intervention and telephone intervention.}
\begin{tabular}{lcccccccccccc}
\hline \hline
&  & \multicolumn{5}{c}{BDPREM} & \multicolumn{1}{c}{} & \multicolumn{5}{c}{PREM} \\
\cline{3-7} \cline{9-13} Parameter & Truth & MSE & Bias & Var & Avg Var & 95\% CP& & MSE & Bias & Var & Avg Var & 95\% CP \\
\hline
Intercept & -0.42 & 0.14 & -0.16 & 0.11 & 0.09 & 0.85& & 2.92 & -1.65 & 0.21 &0.14& 0.02\\
IDU & 0.48 & 0.08 & -0.03 & 0.08 & 0.10& 0.96 & & 0.23 & 0.08 & 0.22 &0.23& 0.89 \\
MSM & 0.60 & 0.08 & -0.04 & 0.08 &0.08& 0.94 & & 0.28 & 0.24 & 0.22 &0.19 & 0.89 \\
CASUAL & 1.16 & 0.06 & 0.07 & 0.05 &0.05 & 0.93 & & 0.08 & 0.04 & 0.08 & 0.01 & 0.34 \\
TRADE & 0.92 & 0.06 & 0.07 & 0.05 &0.06& 0.91 & & 0.12 & 0.04 & 0.11 & 0.01 & 0.31 \\
Month 3 & -0.32 & 0.18 & 0.13 & 0.17 &0.11& 0.83 & & 0.27 & 0.04 & 0.27 & 0.01 & 0.33 \\
Month 6 & -0.60 & 0.19 & 0.20 & 0.15 &0.12& 0.83& & 0.35 & 0.08 & 0.34 & 0.01 & 0.28 \\
Month 9 & -0.87 & 0.21 & 0.23 & 0.16 &0.15& 0.85& & 0.43 & -0.02 & 0.43 & 0.02 & 0.28 \\
Month 15 & -0.60 & 0.19 & 0.20 & 0.15 &0.14& 0.90& & 0.37 & 0.08 & 0.36 & 0.02 & 0.29 \\
I*Month 3 & -0.65 & 0.19 & -0.10 & 0.18 &0.20& 0.92& & 0.53 & -0.01 & 0.53 & 0.04 & 0.38 \\
I*Month 6 & -0.27 & 0.28 & -0.20 & 0.24 &0.21& 0.85& & 0.84 & -0.13 & 0.83 & 0.04 & 0.34 \\
I*Month 9 & 0.35 & 0.28 & -0.24 & 0.22 &0.22& 0.82& & 0.57 & 0.03 & 0.57 & 0.04 & 0.38 \\
I*Month 15 & -0.25 & 0.29 & -0.14 & 0.27 &0.21& 0.86& & 0.66 & -0.06 & 0.65 & 0.04 & 0.36 \\
T*Month 3 & 0.82 & 0.23 & -0.13 & 0.21 &0.17& 0.91& & 0.46 & -0.05 & 0.46 & 0.02 & 0.31 \\
T*Month 6 & 0.48 & 0.33 & -0.28 & 0.26 &0.22& 0.83& & 0.63 & -0.16 & 0.60 & 0.03 & 0.35 \\
T*Month 9 & 1.66 & 0.26 & -0.27 & 0.18 &0.19& 0.84& & 0.67 & 0.03 & 0.67 & 0.03 & 0.34 \\
T*Month 15 & 0.58 & 0.29 & -0.22 & 0.24 &0.21& 0.88& & 0.65 & -0.14 & 0.63 & 0.03 & 0.28 \\ \hline
$D_\beta$ & 0.98 & 0.98 & -0.0049 & 0.98 & 0.06& 0.96& & 34.16 & 5.28 & 6.26 & 1.20 & 0.00 \\ \hline
\hline
\end{tabular}
\label{simulation1}
\end{table}
\end{landscape}
\clearpage

\begin{table}
\centering
\caption{True values of the regression coefficients $\bfpsi$ in the BD process, and mean of the posterior means estimated from the generated data. Mean square error (MSE), bias, variance (Var), average over the variances from the 100 simulated data sets (Avg Var), 95\% confidence coverage proportion (CP) for the regression coefficients $\bfpsi$ in the BD process are presented. PB is post-baseline.  I and T represent indicators of being in the in-person intervention and telephone intervention.}
\begin{tabular}{lccccccc}
\hline \hline
Parameter & True value & Mean & MSE & Bias & Var & Avg Var & 95\% CP \\ \hline
Intercept & 2.0 & 1.94 & 0.05 & -0.06 & 0.04 & 0.04 & 0.93 \\
PB & -0.5 & -0.35 & 0.09 & 0.16 & 0.06 & 0.06 & 0.91\\
PB*I & 0.5 & 0.37 & 0.08 & -0.17 & 0.05 & 0.06 & 0.93 \\
PB*T & -0.5 & -0.57 & 0.08 & -0.12 & 0.07 & 0.07 & 0.92 \\
CASUAL & 0.5 & 0.51 & 0.03 & 0.02 & 0.03 & 0.04 & 0.97 \\
TRADE & 0.5 & 0.56 & 0.05 & 0.07 & 0.05 & 0.05 & 0.96 \\\hline
\end{tabular}
\label{simulation psi}
\end{table}
\clearpage

\begin{landscape}
\begin{table}
\caption{Comparison of mean square error (MSE), bias, variance (Var), average over the variances from the 100 simulated data sets (Avg Var), 95\% confidence coverage percentage (CP) for the regression coefficients $\alpha$ of the PREM component and $D_\beta$ from the BDPREM and the PREM, using data generated from PREM with the same true values of $\alpha$ and $D_\beta$ as in Table {\ref{simulation1}}. MSE, Bias, Var, and Avg Var are multiplied by 100.  I and T represent indicators of being in the in-person intervention and telephone intervention.}
\begin{tabular}{lcccccccccccc}
\hline \hline
& & \multicolumn{5}{c}{PREM with BD} & \multicolumn{1}{c}{} & \multicolumn{5}{c}{PREM} \\
\cline{3-7} \cline{9-13} Parameter & Truth & MSE & Bias & Var & Avg Var & 95\% CP & & MSE & Bias & Var & Avg Var & 95\% CP \\
\hline
Intercept & -0.42 & 4.1 & 1.4 & 4.1 & 2.7 & 88 & & 4.1 & 0.3 & 4.1 & 2.8 & 87 \\
IDU & 0.48 & 6.0 & -0.4 & 6.0 & 4.9 & 94 & & 5.6 & -0.1 & 5.6 & 4.9 & 95 \\
MSM & 0.60 & 5.6 & -2.8 & 5.5 & 3.7 & 88 & & 5.7 & -2.0 & 5.6 & 3.8 & 87 \\
CASUAL & 1.16 & 0.4 & 0.9 & 0.4 & 0.5 & 96 & & 0.4 & 1.1 & 3.9 & 0.5 & 95 \\
TRADE & 0.92 & 0.4 & -0.2 & 0.4 & 0.5 & 96 & & 0.4 & -0.3 & 0.4 & 0.4 & 96 \\
Month 3 & -0.32 & 1.0 & 0.3 & 1.0 & 1.3 & 98 & & 1.0 & -0.02 & 1.0 & 1.2 & 95 \\
Month 6 & -0.60 & 1.4 & -0.8 & 1.4 & 1.3 & 95 & & 1.5 & -1.2 & 1.5 & 1.1 & 91 \\
Month 9 & -0.87 & 1.7 & 1.1 & 1.6  & 1.6  & 94 & & 1.7 & 0.3 & 1.7 & 1.6 & 92 \\
Month 15 & -0.60 & 1.5 & 0.6 & 1.5 & 1.7 & 97 & & 1.5 & 0.3 & 1.5 & 1.5 & 97\\
I*Month 3 & -0.65 & 3.3 & -1.7 & 3.3 & 3.1 & 95 & & 3.4 & -1.3 & 3.4 & 3.1 & 98\\
I*Month 6 & -0.27 & 3.5 & -0.6 & 3.5 & 3.2 & 91 & & 3.7 & -0.4 & 3.7 & 3.0 & 92 \\
I*Month 9 & 0.35 & 3.3 & -3.0 & 3.2 & 3.3 & 94 & & 3.5 & -2.0 & 3.4 & 3.1 & 94\\
I*Month 15 & -0.25 & 4.3 & -0.6 & 4.3 & 3.7 & 95 & & 4.3 & -0.2 & 4.3 & 3.5 & 94 \\
T*Month 3 & 0.82 & 2.0 & -1.6 & 2.0 & 2.0 & 96 & & 1.9 & -0.9 & 1.9 & 1.9 & 93\\
T*Month 6 & 0.48 & 2.7 & -0.3 & 2.7 & 2.4 & 94 & & 3.0 & 0.5 & 3.0 & 2.2 & 88 \\
T*Month 9 & 1.66 & 2.3 & -1.7 & 2.3 & 2.2 & 94 & & 2.5 & -0.4 & 2.5 & 2.1 & 94 \\
T*Month 15 & 0.58 & 2.5 & -1.5 & 2.5 & 2.7 & 96 & & 2.5 & 0.2 & 2.5 & 2.4 & 96 \\ \hline
$D_\beta$ & 0.98 & 94.1 & -4.0 & 94.0 & 1.7 & 97 & & 95.1 & -2.9 & 95.1 & 1.7 & 97 \\ \hline
\hline
\end{tabular}
\label{simulation2}
\end{table}
\end{landscape}
\clearpage

\begin{figure}
\centering
\subfloat[$Z = 2, \lambda = 0.5$]{
\includegraphics[scale=.35]{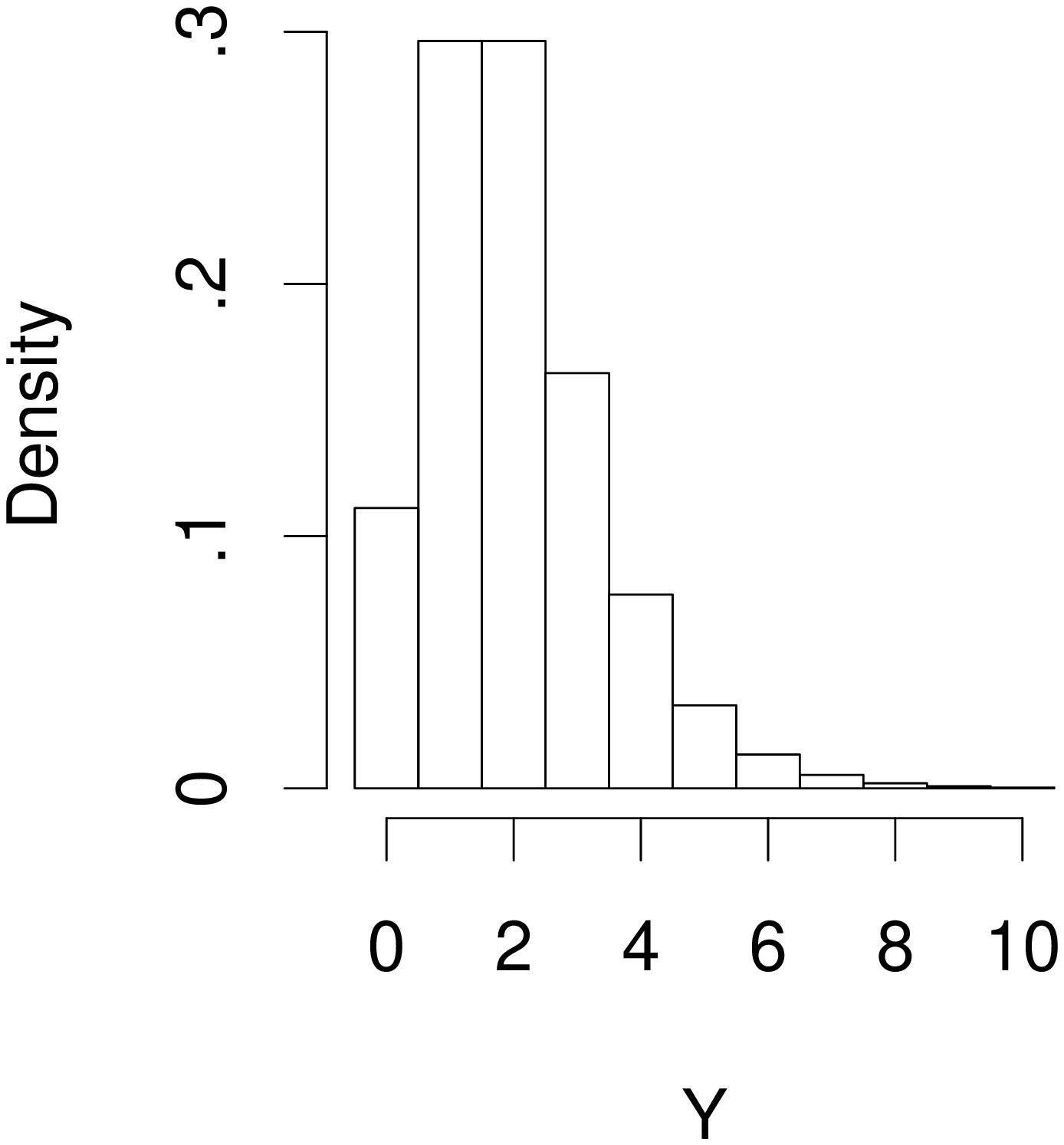}
\label{Ydistri_1}
}
\subfloat[$Z = 15, \lambda = 0.5$]{
\includegraphics[scale=.35]{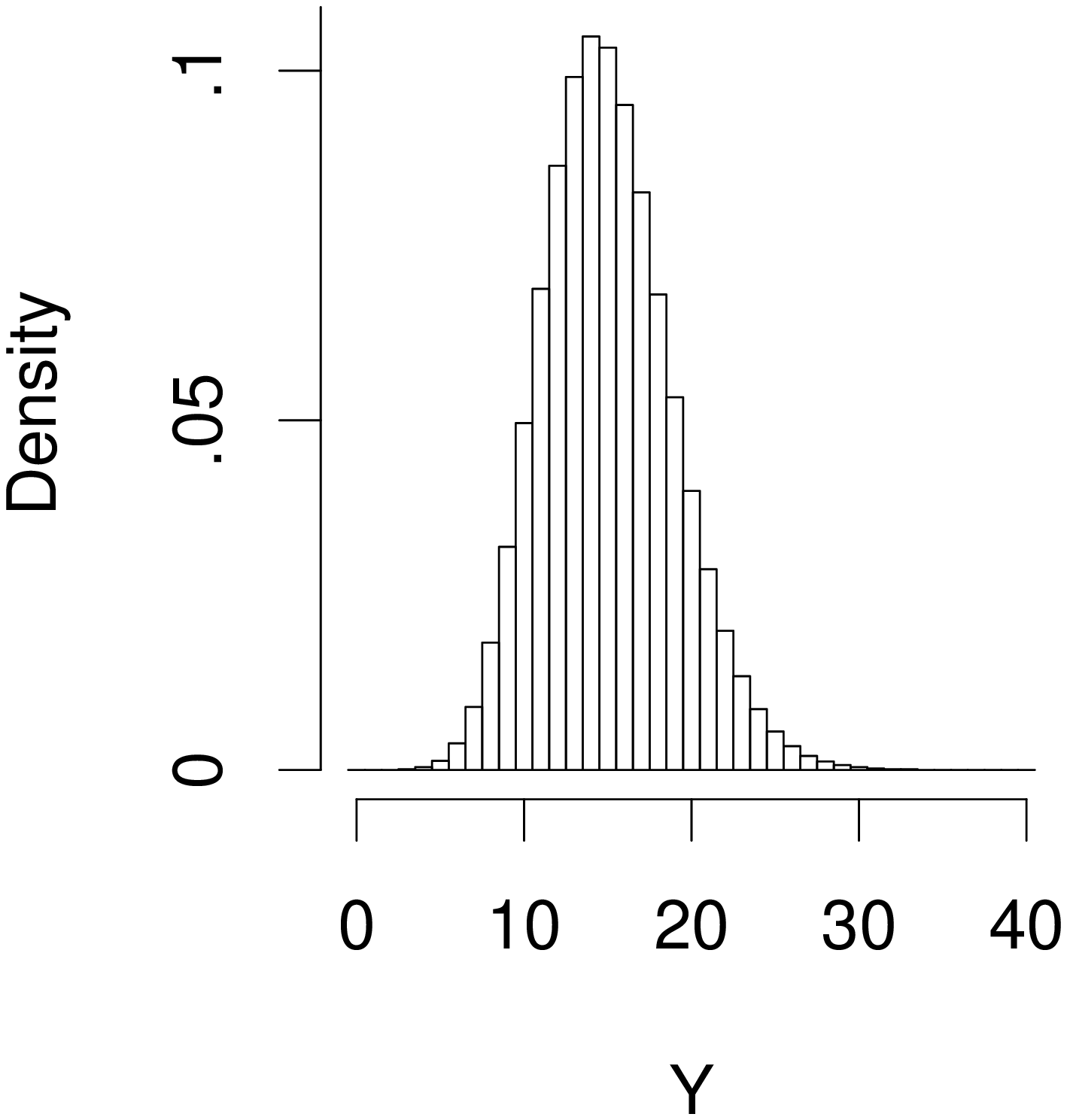}
\label{Ydistri_4}
}
\subfloat[$Z = 50, \lambda = 0.5$]{
\includegraphics[scale=.35]{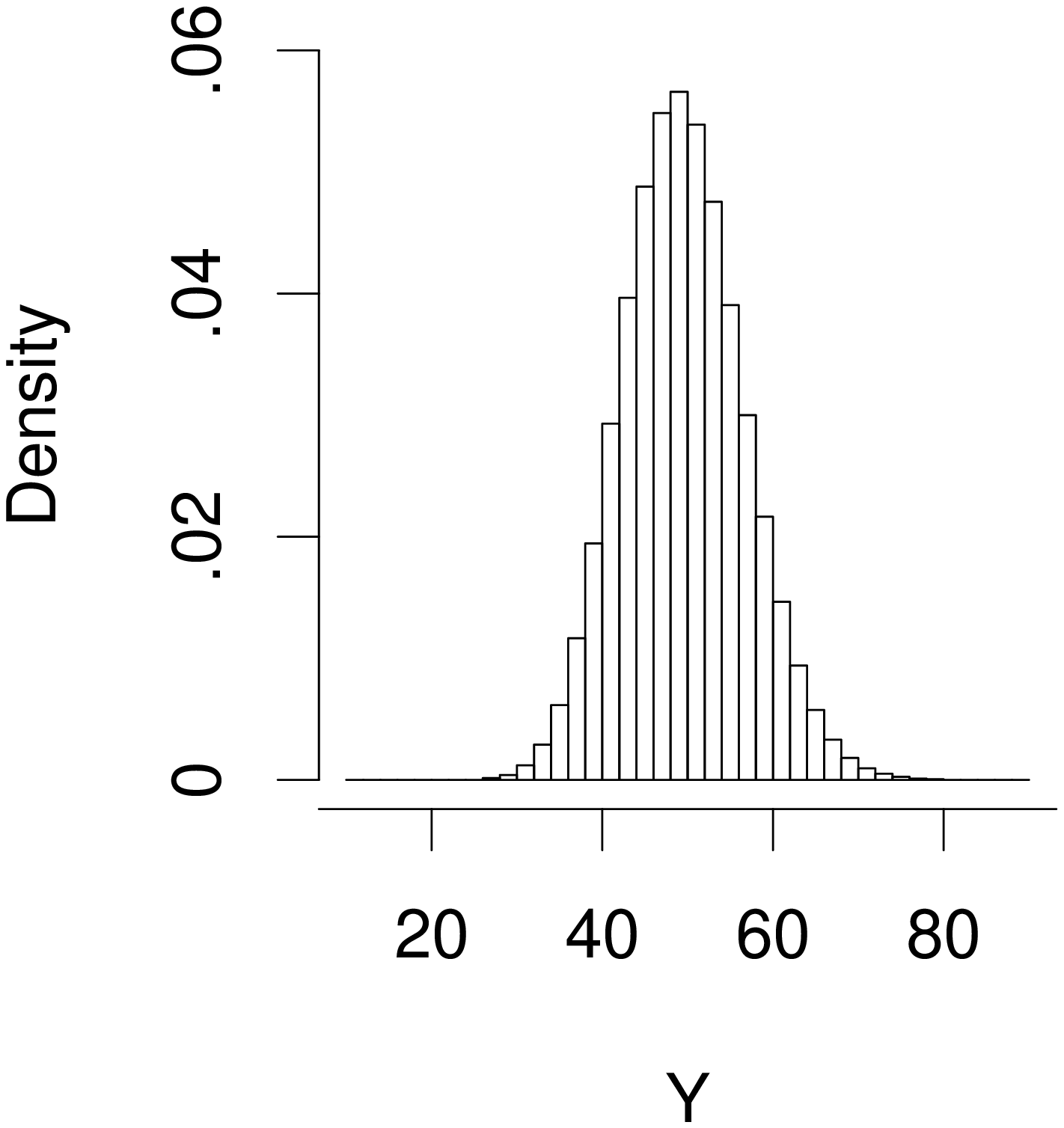}
\label{Ydistri_7}
}\\
\subfloat[$Z = 2, \lambda = 3$]{
\includegraphics[scale=.35]{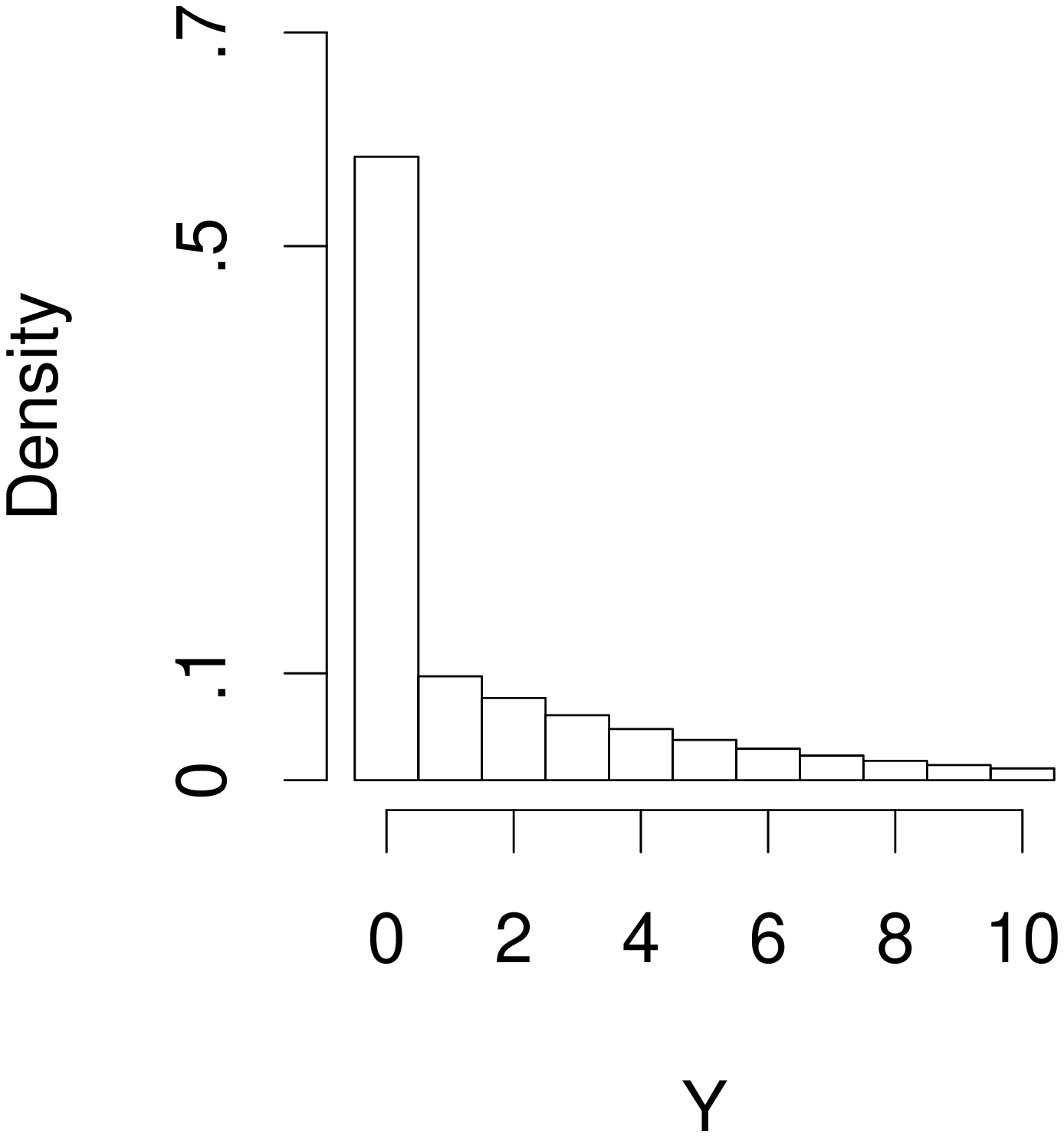}
\label{Ydistri_2}
}
\subfloat[$Z = 15, \lambda = 3$]{
\includegraphics[scale=.35]{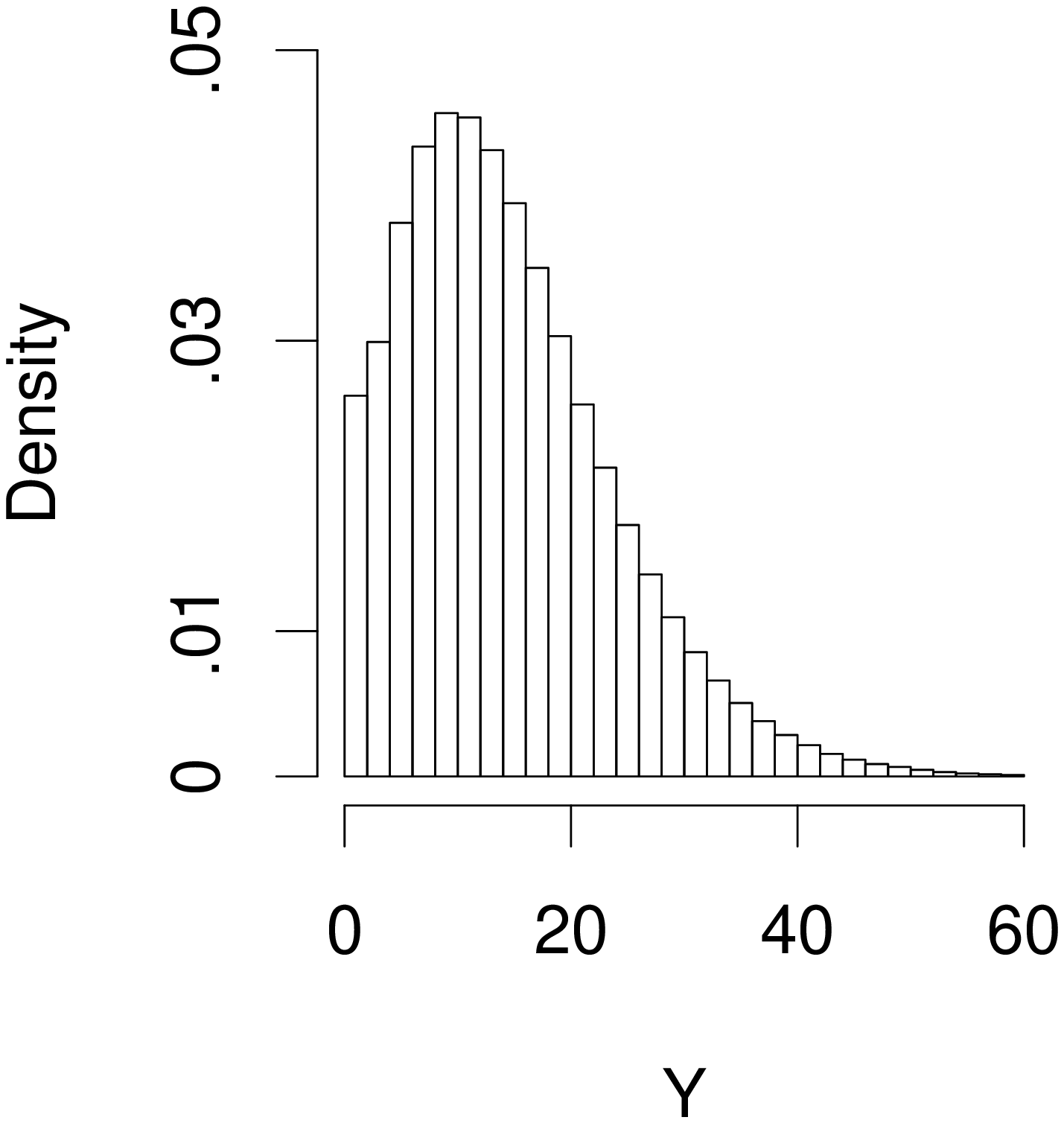}
\label{Ydistri_5}
}
\subfloat[$Z = 50, \lambda = 3$]{
\includegraphics[scale=.35]{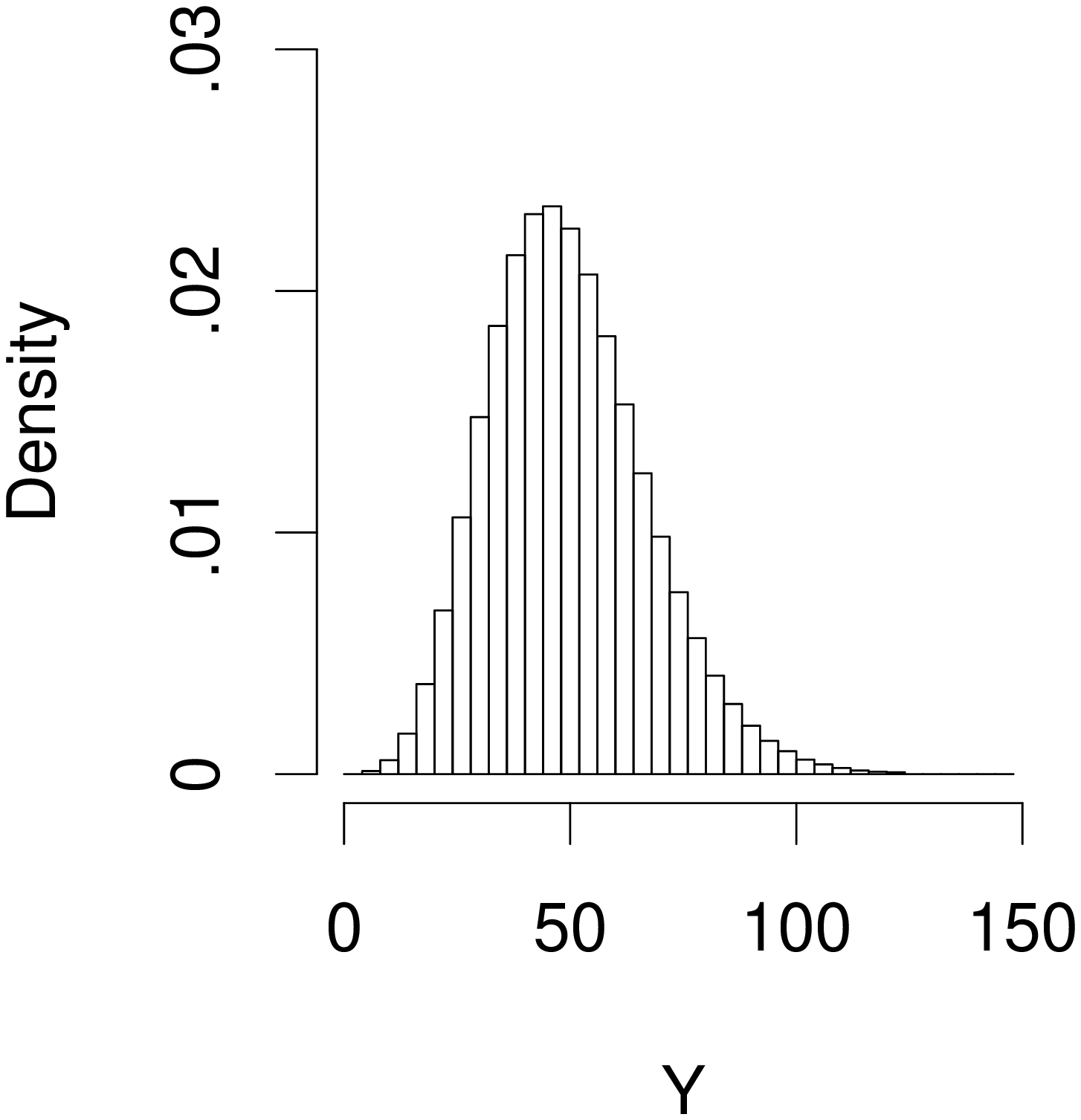}
\label{Ydistri_8}
}\\
\subfloat[$Z = 2, \lambda = 7$]{
\includegraphics[scale=.35]{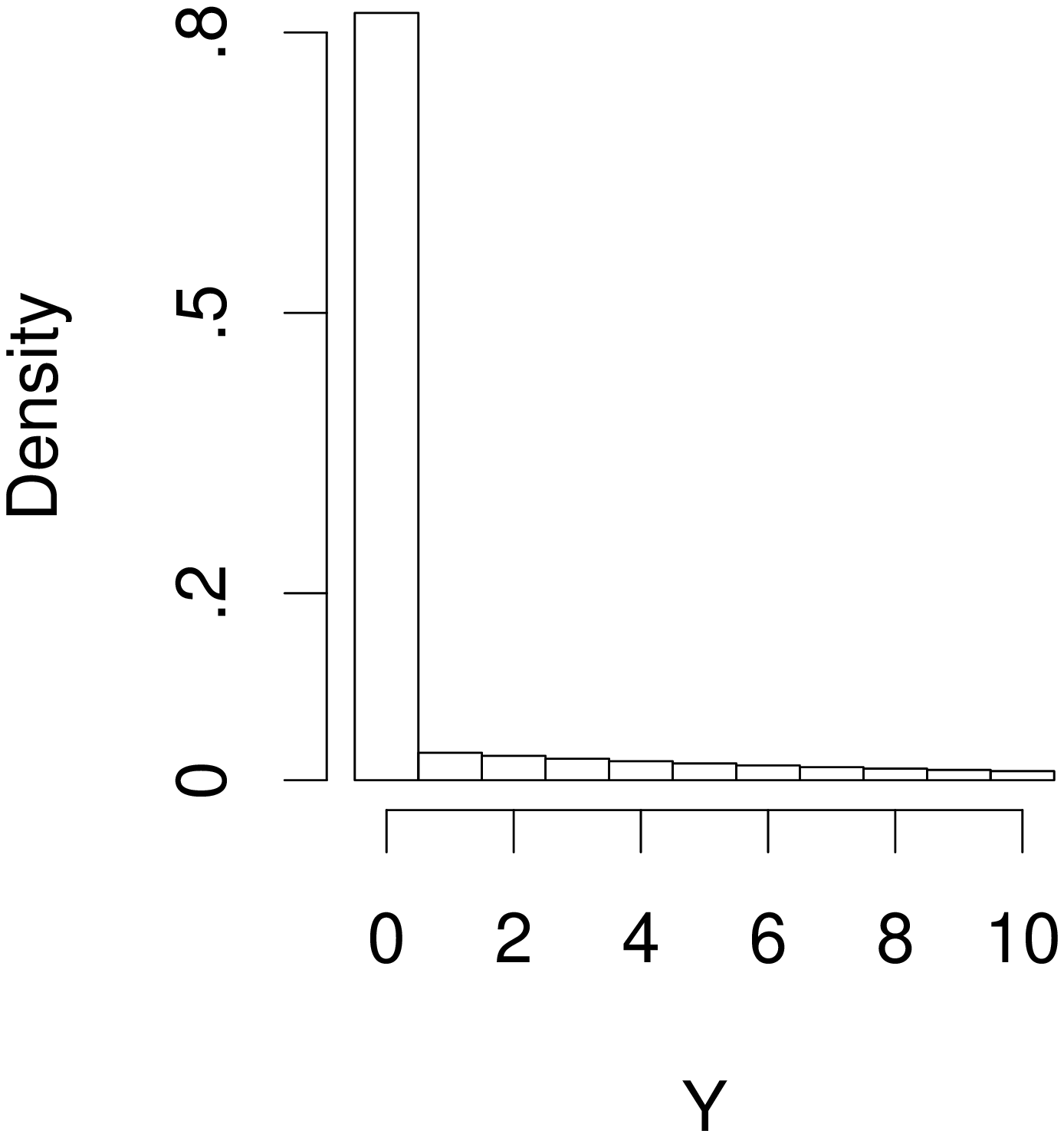}
\label{Ydistri_3}
}
\subfloat[$Z = 15, \lambda = 7$]{
\includegraphics[scale=.35]{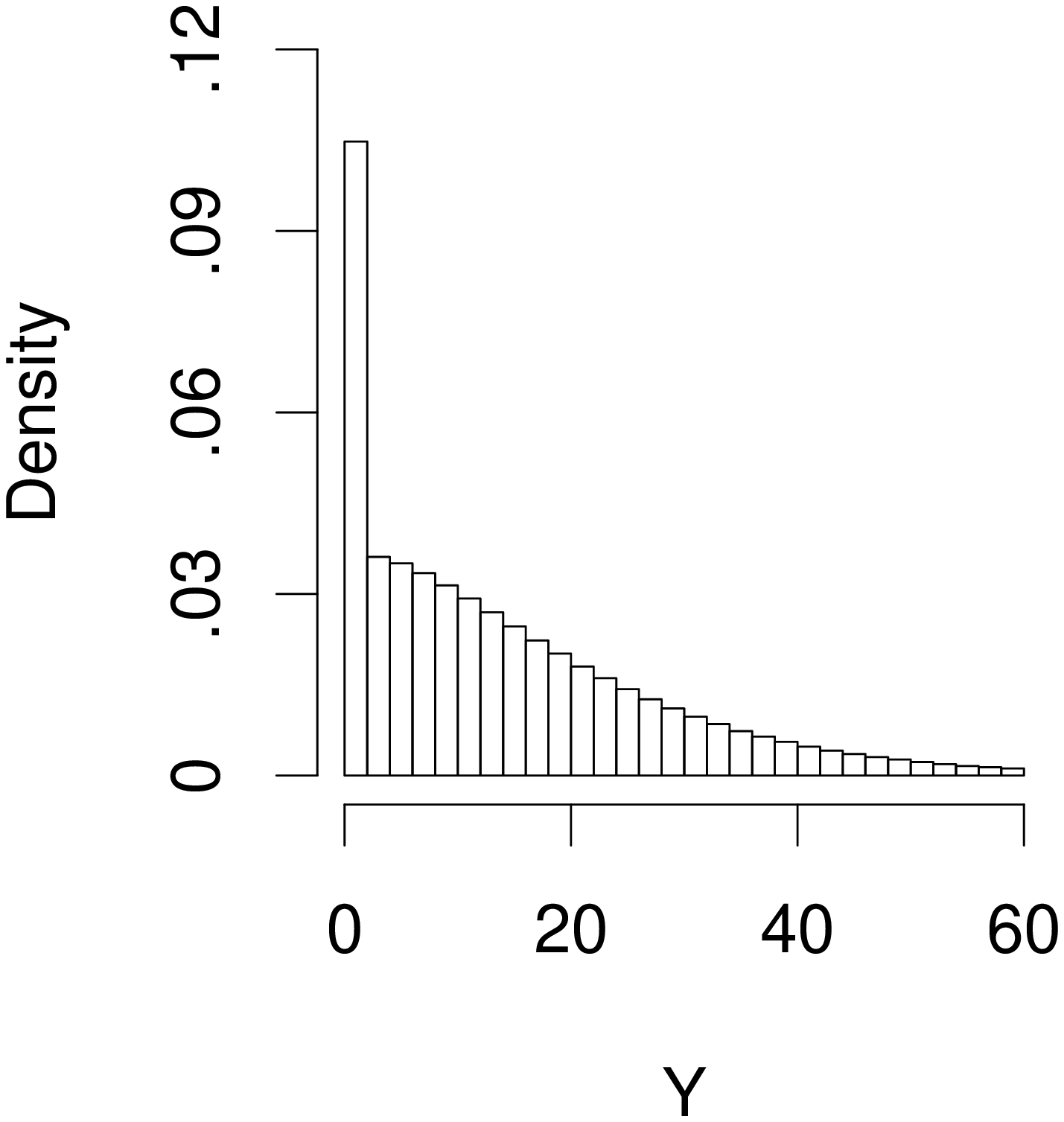}
\label{Ydistri_6}
}
\subfloat[$Z = 50, \lambda = 7$]{
\includegraphics[scale=.35]{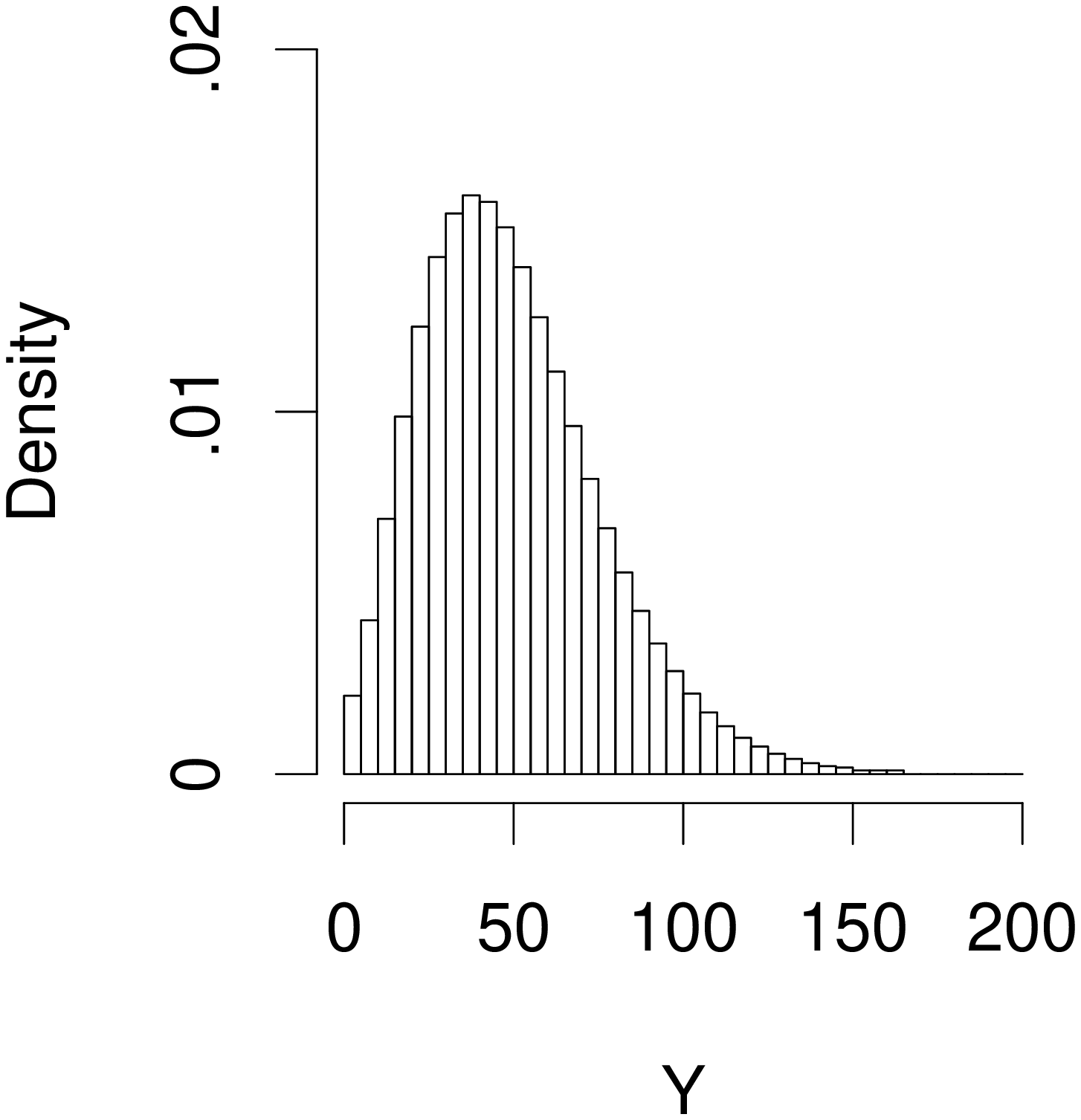}
\label{Ydistri_9}
}
\caption{\label{fig:Ydistri} Conditional distributions of reported count $Y$ given values of true count $Z$ and equal birth and death rate $\lambda$.  Row 1 has small $\lambda = 0.5$ and row 3 has large $\lambda = 7$. Column 1 has small $Z = 2$ and column 3 has large $Z = 50$.}
\end{figure}

\setlength{\oddsidemargin}{15pt}
\clearpage

\clearpage

\end{document}